\documentclass[a4paper]{report}
\usepackage[utf8]{inputenc}
\usepackage[T1]{fontenc}
\usepackage{RJournal}
\usepackage{amsmath,amssymb,array}
\usepackage{booktabs}

\usepackage{longtable}

\makeatletter
 \let\@cite@ofmt\@firstofone
 \def\@biblabel#1{}
 \def\@cite#1#2{{#1\if@tempswa , #2\fi}}
\makeatother
\newlength{\cslhangindent}
\newlength{\csllabelwidth}
 {\begin{list}{}{%
  \setlength{\itemindent}{0pt}
  \setlength{\leftmargin}{0pt}
  \setlength{\parsep}{0pt}
  \ifodd #1
   \setlength{\leftmargin}{\cslhangindent}
   \setlength{\itemindent}{-1\cslhangindent}
  \fi
  \setlength{\itemsep}{#2\baselineskip}}}
 {\end{list}}
\usepackage{calc}

\usepackage{booktabs}
\usepackage{colortbl}
\usepackage{colortbl}
\usepackage{tabu}
\usepackage{xcolor}
\usepackage{amsmath}
\usepackage{tabularx}
\usepackage{array}

\begin{document}

\sectionhead{Contributed research article}
\volume{XX}
\volnumber{YY}
\year{20ZZ}
\month{AAAA}

\begin{article}
\title{changepointGA: An R package for Fast Changepoint Detection via Genetic Algorithms}

\author{by Mo Li and QiQi Lu}

\maketitle

\abstract{%
Detecting changepoints in a time series of length \(N\) entails evaluating up to \(2^{N-1}\) possible changepoint models, making exhaustive enumeration computationally infeasible. Genetic algorithms provide a stochastic way to identify the structural changes: a population of candidate models evolves via selection, crossover, and mutation operators until it converges on one changepoint model that balances the goodness-of-fit with parsimony. The R package changepointGA encodes each candidate model as an integer chromosome vector and supports both the basic single-population model and the island model. Parallel computing is implemented on multi-core hardware to further accelerate computation. Users may supply custom fitness functions or genetic operators, while a streamlined interface supports routine analyses. Extensive simulations demonstrate that our package runs significantly faster than binary-encoded genetic algorithm alternatives. Additionally, this package can simultaneously locate changepoints, estimate their effects, and select time series model orders, such as autoregressive and moving-average orders. Applications to array-based comparative genomic hybridization data and a century-long temperature series further highlight the package's value in biological and climate research.
}

\section{Introduction}\label{introduction}

In statistical analysis and time series modeling, a changepoint refers to a time at which there is a change in the underlying statistical properties or behavior of the data. It could indicate a change in the mean \citep{robbins2011mean, aue2013structural}, variance \citep{gombay1996estimators, aue2013structural, chapman2020nonparametric}, or autocorrelation structure, such as a shift in the lag dependence of the series \citep{davis2006structural, aue2013structural, dette2019change}. Detecting changepoints is crucial across various fields \citep{chen2000parametric} such as finance \citep{andreou2002detecting}, economics \citep{bai1998estimating, zeileis2005monitoring}, environmental science \citep{li2022changepoint}, and signal processing \citep{gavrin2018detection}, as it helps identify shifts in trends, detect anomalies, and make predictions.

Since \citet{page1955test} first explored changepoint detection on independently and identically distributed (IID) data, the literature on this topic has grown significantly. A common strategy is to assume a parametric regression model with potential changepoints in the observed sequence and then compare the fit across different configurations, varying both the number and the locations of changepoints. The best-fitting model is then used to estimate the changepoints \citep{aue2013structural, shi2022changepoint}. However, obtaining this model requires fitting all candidate models across every possible changepoint configuration, which is computationally demanding, particularly for long time series. For instance, given a time series with \(N\) observations and assuming at most one changepoint (AMOC), there are \(N-1\) candidate configurations, each requiring a separate model fit. When multiple changepoints are allowed, the number of candidate configurations increases exponentially to \(2^{N-1}\) \citep{lee2020trend}. This exponential growth of the search space renders the problem computationally intractable, and consequently, exhaustive search becomes impractical even for moderate values of \(N\).

Rather than exhaustively searching all possible changepoint models to identify the globally optimal model, genetic algorithms aim to efficiently explore the configuration space and find a solution that provides the optimal fit. Originally introduced by \citet{holland1975adaptation} as a stochastic search technique, genetic algorithms have proven effective for solving optimization problems involving a dynamically changing number of parameters. They operate by mimicking the evolutionary process in nature, progressively refining candidate solutions across successive generations. The \CRANpkg{GA} optimization package in R \citep{R}, developed by \citet{scrucca2013ga}, serves as a general-purpose framework for implementing genetic algorithms in search spaces comprising both continuous- and discrete-valued solutions. Improvements such as the island model genetic algorithm and parallel computation were later incorporated into the \CRANpkg{GA} package \citep{scrucca2017ga}. Within this framework, a common approach to representing a changepoint configuration is to encode it as a binary string of length \(N\) \citep{shi2022changepoint}. In this representation, a value of one in the binary string indicates the presence of a changepoint at the corresponding time point, whereas a value of zero indicates its absence. However, this encoding method has three notable limitations in the context of changepoint detection.

First, although genetic algorithms with binary encoding reduce the need to fit every possible changepoint model, it represents candidate solutions as vectors of length \(N\), resulting in high-dimensional chromosomes, especially for long time series. Second, binary encoding permits unrealistic changepoint configurations, such as placing a changepoint at \(t=1\) or assigning two adjacent changepoints at times \(t\) and \(t+1\). Such configurations can lead to convergence issues during parameter estimation (e.g., computational singularities when inverting matrices), particularly in models with complex structures. Third, for time series models with changepoints, such as piecewise autoregressive moving-average (ARMA) processes \citep{davis2006structural,shi2022changepoint}, piecewise trend-stationary series with periodic autoregressive (PAR) innovations \citep{lu2010mdl}, the \CRANpkg{GA} package requires users to pre-specify order parameters (i.e., AR, MA, or PAR orders) to run the algorithm \citep{shibata1976selection, scrucca2017ga}. These parameters define the number of past observations or residuals used to model temporal dependencies. When the true orders are unknown, a separate model selection step is required to estimate them. However, standard estimation methods often degrade unless the changepoint locations are known a priori. Ideally, a changepoint detection method should dynamically and simultaneously estimate both the model structure and the changepoint locations to enable more robust analysis.

Instead of using a purely binary chromosome representation, \citet{davis2006structural} used a length-\(N\) encoding in which non-breakpoint locations are marked by \(-1\), whereas segment-start locations store the AR order of the corresponding segment. This representation enables the genetic algorithm to estimate changepoint locations and segment-specific AR orders simultaneously, but the chromosome length remains tied to the length of the time series. By contrast, \citet{lu2010mdl} adopted a more compact representation in which the number of changepoints and their locations are encoded directly as integer-valued vectors. This approach substantially reduces chromosome dimensionality and improves both computational and memory efficiency. Related genetic-algorithm encodings have also been applied to changepoint detection in subsequent work \citep{polushina2011change, lee2014trends, li2026MTMcpt}.

Building on this idea, \CRANpkg{changepointGA} provides a specialized R framework for efficient changepoint detection in time series using genetic algorithms with a compact integer-encoded chromosome representation. In contrast to the general-purpose stochastic search framework provided by \CRANpkg{GA}, \CRANpkg{changepointGA} is designed specifically for changepoint analysis. Its main practical advantages include substantially reducing computational cost, enforcing a minimum segment-length constraint to exclude unrealistic nearby changepoints, and allowing time series model orders to be estimated jointly with changepoint locations.

The \CRANpkg{changepointGA} package is publicly available from the Comprehensive R Archive Network (CRAN). In genetic algorithms, the fitness or objective function is typically problem-dependent. In the changepoint detection literature, commonly used objective functions include the Bayesian information criterion (BIC) \citep{yao1988estimating}, modified BIC (mBIC) \citep{zhang2007modified}, and minimum description length (MDL) \citep{davis2006structural}. Accordingly, \CRANpkg{changepointGA} provides a flexible implementation of genetic algorithms that allows users to supply problem-specific fitness functions and to define customized genetic operators using either R or \CRANpkg{Rcpp}. Although the algorithm is executed sequentially by default, the package also supports parallel computation to improve performance, particularly when fitness evaluations are computationally intensive.

Furthermore, \CRANpkg{changepointGA} can be applied to a wide range of time series settings, including piecewise stationary ARMA processes and piecewise trend-stationary series. However, care is needed for integrated processes, such as random walks with drift or deterministic trends, because stochastic trends induced by unit roots may be mistaken for level shifts in finite samples. As a preliminary diagnostic step, users may consider conducting appropriate unit root tests before applying \CRANpkg{changepointGA} \citep{perron1989great, zivot1992further}.

The remainder of the paper is organized as follows. Section 2 provides a structured overview of genetic algorithms for changepoint detection. Section 3 introduces the main functions \code{cptga} and \code{cptgaisl} in \CRANpkg{changepointGA}. Section 4 presents simulation studies for changepoint detection with and without ARMA order selection. Two real-world case studies are included in Section 5 to highlight the utility of the proposed framework. Finally, Section 6 summarizes the main contributions and offers concluding remarks.

\section{Background}\label{background}

For any pre-specified time series model with a given set of changepoint locations, model fit is evaluated using a fitness function \(Q(\boldsymbol{\theta})\), where \(\boldsymbol{\theta}=(\boldsymbol{s},\boldsymbol{\tau},\boldsymbol{\beta})'\) denotes the full parameter vector. Here, \(\boldsymbol{s}\) represents the set of model hyperparameters, which may include the AR order, the MA order, or the periodic AR order in PAR models. The vector \(\boldsymbol{\tau}=\{\tau_{1}, \ldots, \tau_{m}\}\) specifies the changepoint locations such that \(1\leq\tau_{1}<\tau_{2}<\cdots<\tau_{m}< N\), with the number of changepoints \(m\) inferred as part of the estimation process. Once \(\boldsymbol{s}\) and \(\boldsymbol{\tau}\) are determined, the remaining model parameters in \(\boldsymbol{\beta}\) can be estimated through standard model fitting procedures. Within this framework, changepoint detection is formulated as a combinatorial optimization problem: the goal is to minimize the scalar-valued objective function \(Q(\boldsymbol{\theta})\) and thereby identify the optimal solution
\begin{equation}
    \boldsymbol{\theta}^{*}\equiv 
    \arg \min_{\boldsymbol{\theta}\in\mathbb{S}}Q(\boldsymbol{\theta}) = \{\boldsymbol{\theta}^{*}\in\mathbb{S}:Q(\boldsymbol{\theta}^{*}) \leq Q(\boldsymbol{\theta}), \ \forall\boldsymbol{\theta}\in\mathbb{S}\}, \label{eq:problem}
\end{equation}
where \(\mathbb{S}\) denotes the feasible parameter space. Note that while the problem statement in Equation \eqref{eq:problem} revolves around minimizing \(Q(\boldsymbol{\theta})\), it's trivial to transform the maximization objective into a minimization problem by simply adjusting the sign of the objective function. When estimating the optimal changepoint configuration along with the model orders and parameters, the dimensionality of \(\boldsymbol{\theta}\) varies and depends on the model order and the number of changepoints. This dynamic dimensionality adds further complexity to the challenge of changepoint detection.

The genetic algorithm operates as a stochastic search optimization method inspired by the principle of survival of the fittest individual in ecology. Each individual in the population is characterized by a chromosome that encodes a potential solution. These chromosomes contain values from \(\boldsymbol{\theta}\) within the feasible search space \(\mathbb{S}\) for the optimization problem. Subsequently, the fitness of each individual (or chromosome) is assessed based on the objective function value \(Q(\boldsymbol{\theta})\). In our \CRANpkg{changepointGA} package, the chromosome representation is specified as a vector,
\begin{equation}
    C = (m, \boldsymbol{s}, \boldsymbol{\tau}, N+1)'. \label{eq:chromosome}
\end{equation}
Note that if no model order selection is desired, then \(\boldsymbol{s}\) is omitted and the genetic algorithm detects only the changepoints. This flexibility enables the package to accommodate diverse user requirements and problem settings, ensuring the genetic algorithm is precisely tailored to the optimization task. The changepoint locations in \(\boldsymbol{\tau}\) are encoded as integer values between 1 and \(N-1\), allowing the length of \(\boldsymbol{\tau}\) to vary dynamically with \(m\). The value \(N+1\) is appended at the end of \(C\) to serve as a delimiter marking the boundary of the chromosome. The dimensionality of the model parameter vector \(\boldsymbol{\beta}\) also varies, depending on the values specified in \(\boldsymbol{s}\) and \(\boldsymbol{\tau}\).

\begin{figure}[h!]

{\centering \includegraphics[width=1\linewidth]{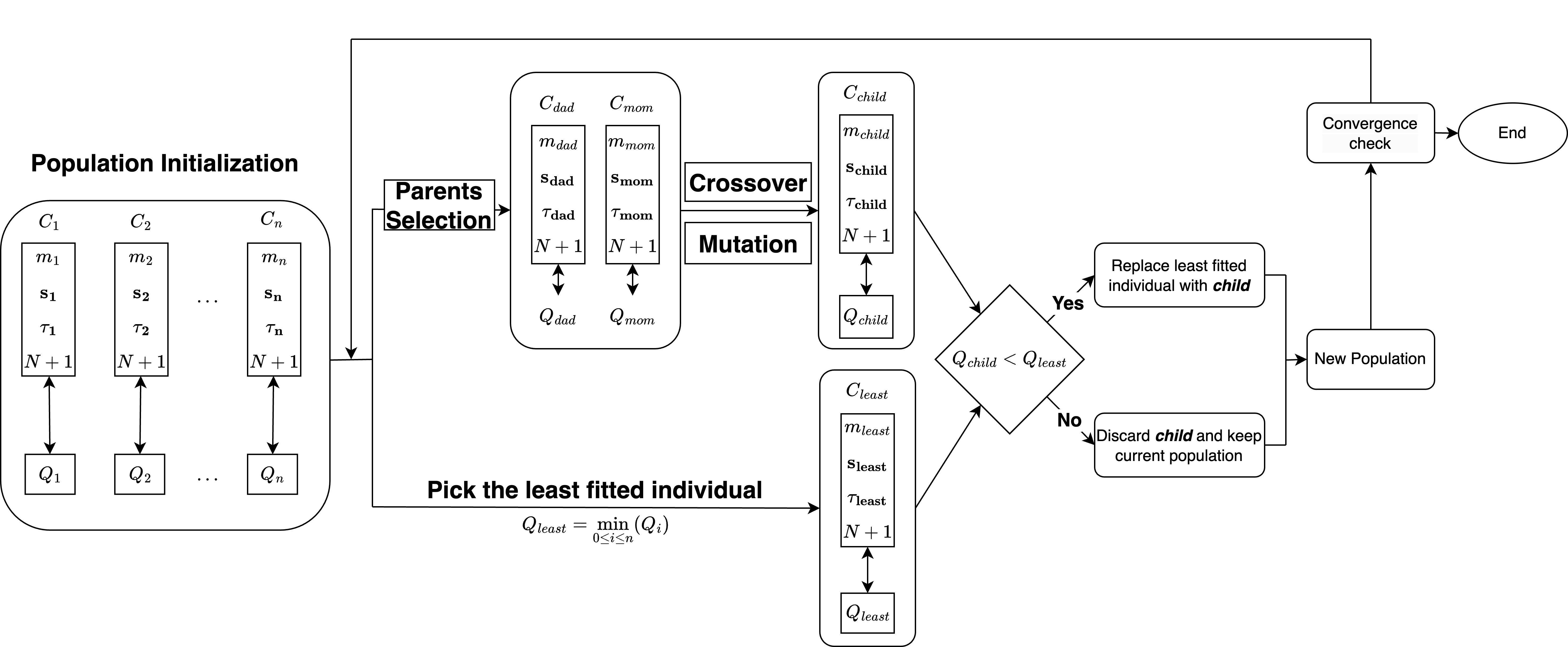} 

}

\caption{The flowchart of the basic genetic algorithm.}\label{fig:GAfig}
\end{figure}

The genetic algorithm starts with the population initialization of size \(n\). For each chromosome (individual), every gene (element) in \(\boldsymbol{s}\) is randomly selected with equal probability from a user-specified candidate set. For example, for an ARMA setting, the AR and MA orders are randomly chosen from the candidate order set, \(\{0, 1, 2, 3, 4\}\), which includes the commonly used values for AR and MA orders. The number of changepoints \(m\) and their locations \(\boldsymbol{\tau}\) are also randomly generated. For each chromosome, the specified model is then fit and its fitness evaluated with respect to the objective function \(Q(\boldsymbol{\theta})\). Subsequently, the selection operator chooses pairs of individuals from the population according to their fitness values, using methods such as the linear ranking algorithm, with higher-fitness individuals (those with smaller \(Q(\boldsymbol{\theta})\)) having a greater probability of being selected as parents. These parents undergo the crossover operator, with a specified crossover probability, to produce offspring (children) for the next generation. Reflecting the biological evolution, each offspring is also subject to a mutation operator that, with a relatively low probability, introduces random genes in its chromosome. This mutation helps the algorithm avoid becoming trapped in local optima. The offspring is then incorporated into the next population using methods such as the steady-state approach from \citet{Davis1991Handbook}. As shown in Figure \ref{fig:GAfig}, if the offspring exhibits superior fitness (i.e., a smaller \(Q(\boldsymbol{\theta})\) than the least fit individual), it replaces that worst individual in the current generation. Otherwise, or if it duplicates an existing chromosome, it is discarded and a new offspring is produced via the selection--crossover--mutation cycle. As generations proceed, the population's best fitness typically improves because fitter individuals are more likely to reproduce. This process repeats until specified termination or convergence criteria are met. The fittest individual in the final population is taken as the approximate optimizer of the objective, and its chromosome yields the estimate of \(\boldsymbol{\theta}\), \(\boldsymbol{\hat{\theta}}\).

\begin{figure}[h!]

{\centering \includegraphics[width=1\linewidth]{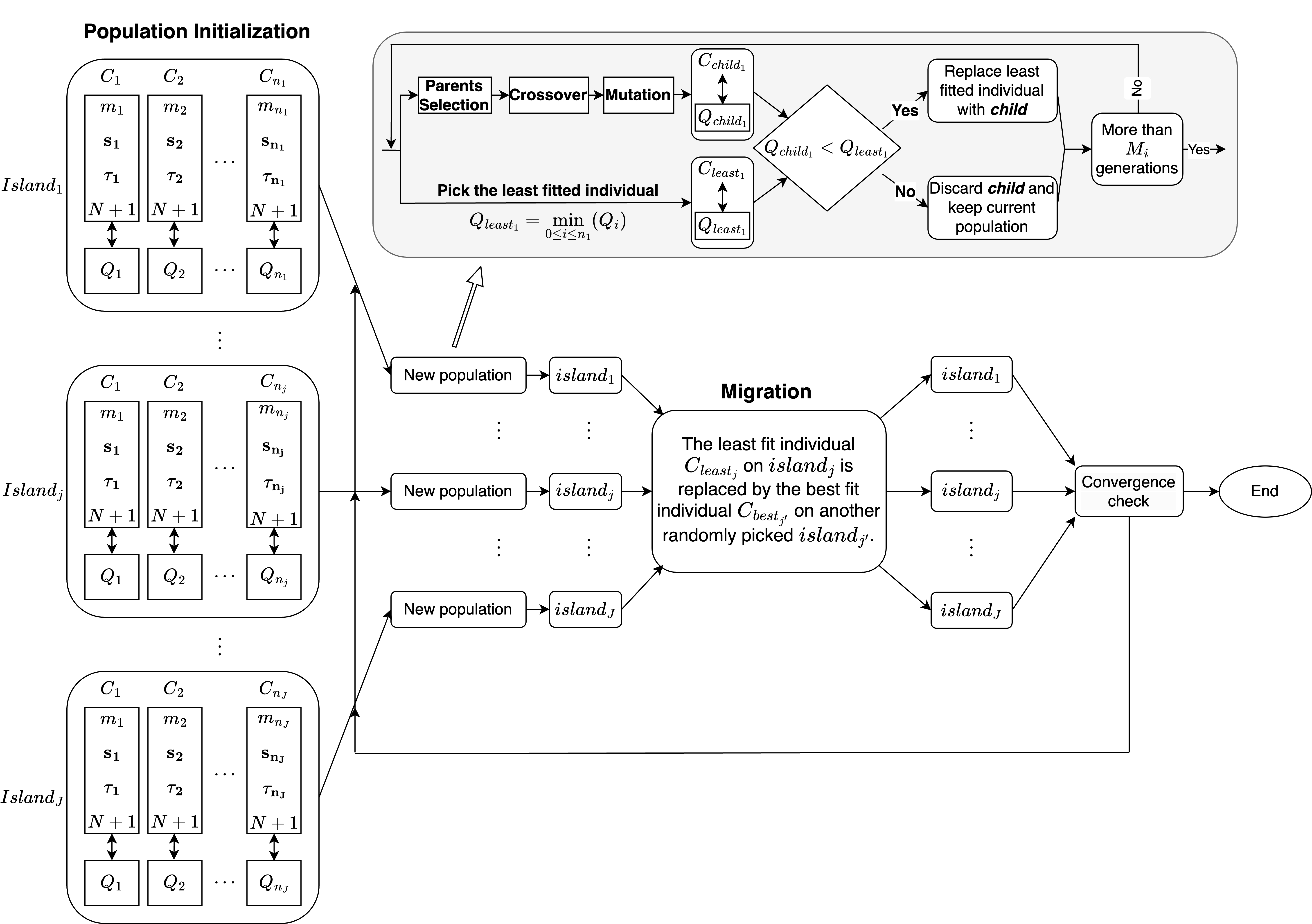} 

}

\caption{The flowchart of the island model genetic algorithm.}\label{fig:IslandGAfig}
\end{figure}

In addition to the standard single-population genetic algorithm, referred to here as the basic genetic algorithm, \CRANpkg{changepointGA} also supports an island model genetic algorithm. The algorithm flowchart is displayed in Figure \ref{fig:IslandGAfig}. Building on the basic genetic algorithm, the initialized population is partitioned into islands (sub-populations). On each island, standard genetic algorithm operators -- parent selection, crossover, mutation, and sub-population replacement -- are executed independently, allowing islands to explore different regions of the search space \(\mathbb{S}\) in parallel. After a predetermined number of generations, individuals from one island migrate to another, enabling the sharing of genetic material between sub-populations. The least fit individual on each island is replaced by the best fit individual from another randomly picked island. This migration maintains genetic diversity and prevents premature convergence to local optima. By balancing isolated evolution with periodic migration, the island model genetic algorithm enhances overall search efficiency and robustness.

\section{\texorpdfstring{The \CRANpkg{changepointGA} package}{The  package}}\label{the-package}

The \CRANpkg{changepointGA} package implements the optimization framework described in Section~2 through two main functions: \code{cptga}, which applies a basic genetic algorithm, and \code{cptgaisl}, which applies an island model genetic algorithm. Both functions take a user-supplied objective function corresponding to $Q(\theta)$ and return an S4 object containing the estimated chromosome, the optimized objective value, and the main genetic algorithm tuning settings. The shared function interface is intended to make it easy to switch between the basic and island model implementations while keeping the same objective function and data inputs. These two primary functions are designed to accommodate the two main optimization settings considered in this paper.

The synopsis of genetic algorithm functions \code{cptga} and \code{cptgaisl} are

\begin{verbatim}
cptga(ObjFunc, N, prange = NULL, option, ...)
cptgaisl(ObjFunc, N, prange = NULL, option, ...)
\end{verbatim}

Both functions share the leading arguments \code{ObjFunc}, \code{N}, and \code{prange}. The argument \code{ObjFunc} specifies the user-defined objective function to be minimized, and \code{N} gives the time series sample size, which is also used to define the terminal chromosome delimiter $N+1$. The argument \code{prange} determines whether model order parameters are included in the chromosome. For changepoint detection only, we set \code{prange=NULL}, and the task is specified by \code{option="cp"}. For joint model order selection and changepoint detection, \code{prange} must be supplied as a list object, and the task is specified by \code{option="both"}. Each element of \code{prange} gives the admissible range for one model order parameter, and the number of model order parameters is determined internally as the length of \code{prange} list. The remaining arguments control the genetic algorithm search, including the population size, crossover and mutation probabilities, changepoint initialization probability, minimum spacing between changepoints, stopping rules, monitoring, parallel computation, and optional user-defined genetic operators. For \code{cptgaisl}, additional island model settings specify the number of islands and the migration-based stopping mechanism.

The argument \code{ObjFunc} must be supplied as an R function that evaluates the fitness of a candidate chromosome and returns a scalar numerical value to be minimized. For example, the \CRANpkg{changepointGA} package includes the BIC-based objective functions for AR(1) model as

\begin{verbatim}
arima_bic(chromosome, plen = 0, XMat, Xt)
\end{verbatim}

where \code{chromosome} is the candidate chromosome to be evaluated, \code{plen} denotes the number of model order parameters encoded in the chromosome, and \code{XMat} and \code{Xt} are model-specific inputs passed from the main genetic algorithm function call. The genetic algorithm functions determine the value of \code{plen} from \code{prange}: \code{plen}=0 when \code{prange=NULL}, and \code{plen=length(prange)} when \code{prange} is supplied as a list. Thus, users should include \code{plen} as an argument in \code{ObjFunc} when the objective function needs to separate the model-order component $s$ from the changepoint component $\tau$ in the chromosome. When \code{prange=NULL}, the chromosome is interpreted as $(m, \boldsymbol{\tau}, N+1)'$; when \code{prange} is a list, the chromosome is interpreted as $(m, \boldsymbol{s}, \boldsymbol{\tau}, N+1)'$. Additional named quantities supplied through \code{...}, such as the observed series, design matrix, or other model-specific constants, are forwarded to \code{ObjFunc} when their names match the arguments of the objective function. If the returned objective value is non-finite, it is treated as \code{Inf}, so failed or inadmissible model fits are penalized during the minimization.

To simplify package usage, the task mode is determined by \code{prange} and \code{option}, while the objective function argument \code{plen} records the number of model order parameters encoded in the chromosome. Table~\ref{tab:taskmode} summarizes the recommended settings for the two main use cases.

\textcolor{blue}{
\begin{table}[ht]
\centering
\caption{Recommended argument settings for the main task modes in \CRANpkg{changepointGA}.}
\label{tab:taskmode}
\begin{tabularx}{\textwidth}{>{\raggedright\arraybackslash}X >{\raggedright\arraybackslash}p{0.22\textwidth} >{\raggedright\arraybackslash}p{0.22\textwidth} c}
\hline
Task & \code{plen} & \code{prange} & \code{option} \\
\hline
Changepoint detection only & 0 & \code{NULL} & \code{"cp"} \\
\hline
Joint model order selection and changepoint detection & number of model order parameters & list of candidate ranges & \code{"both"} \\
\hline
\end{tabularx}
\end{table}
}

The main genetic algorithm tuning parameters control the balance between search breadth, stochastic variation, convergence behavior, and the realism of proposed segmentations. Larger values of population size, \code {popSize}, and, in the island model, the number of islands, \code{numIslands}, broaden the search and are often helpful when changepoint detection is combined with model-order selection. The crossover and mutation probabilities, \code{pcrossover} and \code{pmutation}, regulate the balance between inheriting information from promising chromosomes and introducing new candidate solutions. The parameter \code{pchangepoint} influences how readily chromosomes include changepoints, while \code{minDist} imposes a minimum separation between adjacent changepoints, thereby excluding scientifically implausible short segments. In practice, \code{minDist} should be chosen using domain knowledge about the shortest meaningful regime length. While larger values of \code{minDist} can help improve computational efficiency by pruning the search space and reducing spurious detections, the value must remain smaller than the shortest anticipated true regime to ensure that valid changepoint configurations are not excluded from the feasible set.

The function \code{cptga} is useful when a single evolving population is sufficient, whereas \code{cptgaisl} is preferable when a more robust and efficient search is desired through multiple sub-populations and migration. In the island model, the total population is partitioned into \code{numIslands} islands, each of which evolves independently for a prescribed number of generations, \code{maxgen}, before exchanging highly fit individuals. This strategy often improves exploration of the solution space and can reduce premature convergence.

The package further provides default routines for population initialization, parent selection, crossover, and mutation, while allowing users to replace these components with customized operators written in either R or \CRANpkg{Rcpp}. This flexibility is important when scientific knowledge suggests problem-specific constraints or when more efficient objective function evaluation is needed. In the simulation studies below, we use the default operators to illustrate the standard workflow. Supplementary Section S1 provides an additional example showing how users may define customized objective functions and operators.

For typical analyses, the workflow is straightforward: specify an objective function, choose whether the task is changepoint detection only or joint order selection, define the associated candidate ranges if needed, and then call either \code{cptga} or \code{cptgaisl}. The fitted S4 object can then be examined using \code{summary} and visualized using \code{plot}, while \code{cpt\_dist} can be used in simulation settings to compare estimated and true changepoint configurations. Detailed descriptions of all arguments and package options are provided in the package documentation.

\section{Simulation Studies}\label{simulation-studies}

This section presents illustrative examples of changepoint detection using time series generated from an ARMA\((p,q)\). The same framework readily extends to other time series models by specifying an appropriate fitness function and chromosome encoding using the \CRANpkg{changepointGA}. For the examples below, let \(\{X_t\}_{t=1}^{N}\) denote a series with \(m\) changepoints at unknown locations \(\{\tau_{1}, \ldots, \tau_{m}\}\). These changepoints partition the series into \(m+1\) regimes, where the regimes may differ in mean, trend, or both.

The model considered here is
\begin{equation}
    X_t = \kappa_t + \epsilon_t, \label{eq:timeseriesmodel}
\end{equation}
where \(\epsilon_t\) follows an ARMA(\(p,q\)) process and \(\kappa_t\) denotes the mean function of \(X_t\). Specifically,
\begin{equation}
    \kappa_t =
    \begin{cases}
        \beta_0 + \alpha_0 t, 
        & 1 \leq t \leq \tau_1, \\
        \beta_0 + \alpha_0 t + \Delta_1 + \alpha_1 t, 
        & \tau_1 + 1 \leq t \leq \tau_2, \\
        \quad \vdots \\
        \beta_0 + \alpha_0 t + \Delta_m + \alpha_m t, 
        & \tau_m + 1 \leq t \leq N.
    \end{cases}
    \label{eq:kappa}
\end{equation}
Here, \(\beta_0\) and \(\alpha_0\) are the baseline intercept and trend parameters, respectively, while
\(\Delta_1,\ldots,\Delta_m\) and \(\alpha_1,\ldots,\alpha_m\) denote the segment-specific level and trend changes associated with the changepoints.

When \(m>1\), the goal is to detect multiple changepoints. We first consider settings without model-order selection to demonstrate the use of \CRANpkg{changepointGA} and to compare its detection performance with the general-purpose \CRANpkg{GA} package \citep{scrucca2013ga}, with particular emphasis on computational efficiency. We then present an example in which changepoint detection is performed jointly with model-order selection, a feature not available in \CRANpkg{GA}. In these experiments, the classical BIC is used as an illustrative model-selection criterion. However, the resulting accuracy may depend on the criterion chosen by the user, such as mBIC, MDL, or related alternatives.  An additional AMOC example (\(m=1\)) is provided in the supplementary materials, where we also illustrate how users can define customized objective functions and genetic operators.

\subsection{\texorpdfstring{4.1 ARMA(\(p\), \(q\)) orders known}{4.1 ARMA(p, q) orders known}}\label{armap-q-orders-known}

By default, \CRANpkg{changepointGA} targets multiple changepoint detection. Methods for this task are commonly grouped into two main classes: (i) binary segmentation based on AMOC procedures, and (ii) direct model selection approaches that jointly fit all subsegments. When model orders are known, the latter treats the full series simultaneously by optimizing an objective function that comprises the negative log-likelihood and a penalty that balances goodness of fit and model complexity. For our illustrative example governed by Equation \eqref{eq:timeseriesmodel},  the BIC is defined as:
\begin{equation}
    Q(\boldsymbol{\theta}) = -2\text{log}(L(\boldsymbol{\theta}))+k\text{log}(N),  \label{eq:MultipleBIC}
\end{equation}
where \(k\) denotes the number of model parameters. The term \(-2\text{log}(L(\boldsymbol{\theta}))\) measures the lack of fit for a model with \(m\) changepoints in \(\boldsymbol{\tau}\), while the second term serves as a penalty to prevent overfitting. A smaller BIC indicates a better model fit. Minimizing \(Q(\boldsymbol{\theta})\) selects the best-fitting model, and the resulting changepoint configuration provides the estimated locations of the changepoints.

Assuming \(\{\epsilon_{t}\}\) in Equation \eqref{eq:timeseriesmodel} follows an AR(1) process (\(p=1, q=0\)),
\begin{equation}
\epsilon_{t} = \phi\epsilon_{t-1} + e_{t}, \label{eq:ar1}
\end{equation}
where \(|\phi| < 1\) is the autoregressive parameter and \(\{e_t\}\) denotes a white noise process with mean zero and variance \(\sigma^2\). The corresponding BIC can be computed using the \texttt{arima\_bic} function from the \CRANpkg{changepointGA} package.

In designing the objective function, \texttt{chromosome} argument holds the individual changepoint configuration \(C\) as described in Equation \eqref{eq:chromosome}. First, we extract the number and locations of the changepoints from chromosome for fitness calculation. To inform the package that the model order selection is not considered here, it is essential to set \texttt{plen\ =\ 0}, indicating that the vector including model orders has a length of zero. The time series data in \texttt{Xt} is input into the objective function and will be included as an additional argument passed to both the \texttt{cptga} and \texttt{cptgaisl} functions during execution.

To illustrate the detection of changepoints with our package, we simulated a time series following the models specified in Equations \eqref{eq:timeseriesmodel}, \eqref{eq:kappa}, and \eqref{eq:ar1} with \(N = 1000\), \(\phi = 0.5\), and \(\sigma^2 = 1\).  For simplicity, all simulations below set \(\alpha_i=0\) for \(i=0,1,\ldots,m\), so that the regimes differ only in their mean levels. Trend shifts are illustrated using the real-data application in the next section.

\subsubsection{4.1.1 Two changepoints}\label{two-changepoints}

Two changepoints were introduced at \(\tau_1 = 250\) and \(\tau_2 = 750\), resulting in three segments with means \(\beta_{0} = 0.5, \beta_0+\Delta_1 = 0.5+2 = 2.5\) and \(\beta_0+\Delta_2 = 0.5+(-2) = -1.5\), respectively. The series can be generated using the function \code{ts\_sim}.

\begin{verbatim}
N <- 1000
XMatT <- matrix(1, nrow = N, ncol = 1)
Xt <- ts_sim(Ts = N, beta = 0.5, XMat = XMatT, sigma = 1, phi = 0.5, theta = NULL, 
  Delta = c(2, -2), CpLoc = c(250, 750), seed = 12345)
\end{verbatim}

\begin{figure}[h!]

{\centering \includegraphics[width=0.7\linewidth]{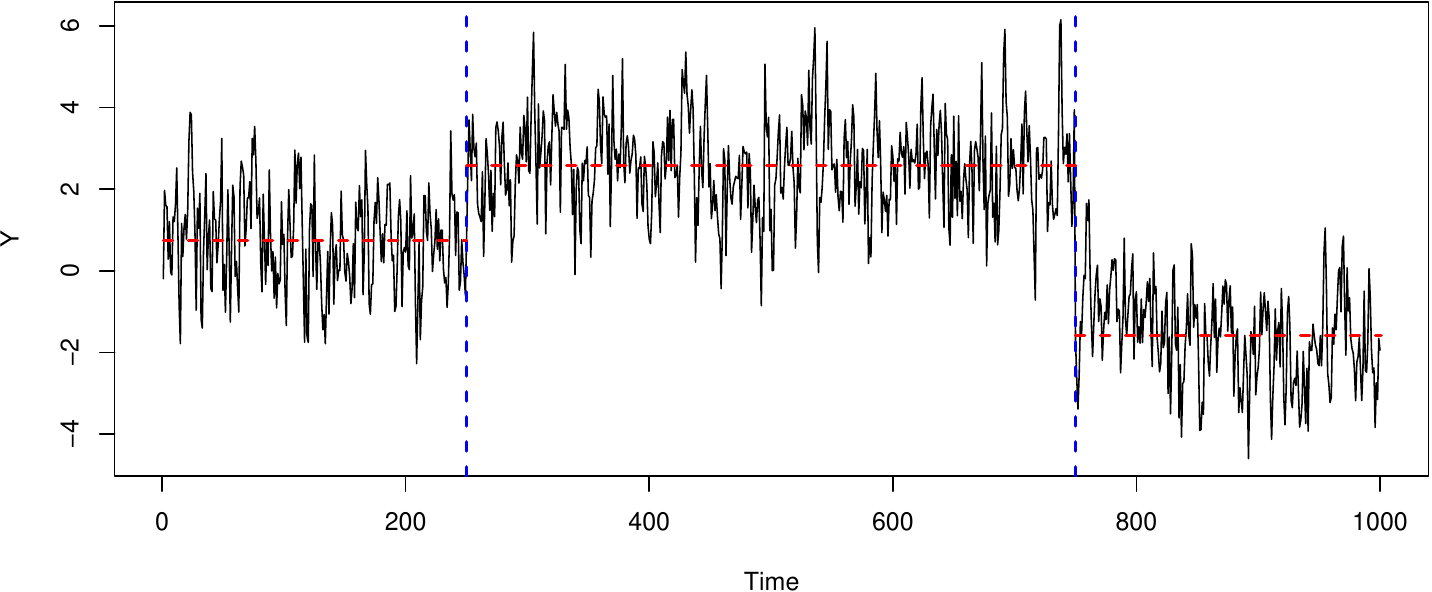} 

}

\caption{A simulated AR(1) time series plot with the changepoint segmentation (blue dashed lines) and the corresponding sample means (red dashed lines).}\label{fig:TsPlotNoOrder}
\end{figure}

\noindent The introduced changepoints can be examined in Figure \ref{fig:TsPlotNoOrder}. The segments are marked by vertical blue dashed lines and the segment sample means are marked by horizontal red dashed lines.

The basic genetic algorithm with the \code{arima\_bic} objective function and default genetic operators can be implemented as follows. During the search, setting \texttt{monitoring\ =\ TRUE} displays the best-fitting changepoint configuration and its corresponding objective function value at each generation. Upon completion, the algorithm returns an S4 object, \texttt{res\_cptga}, containing details of the genetic algorithm hyperparameters and the estimated number and locations of changepoints. A summary can be obtained with the \texttt{summary()} function. The objective function value \(Q(\boldsymbol{\hat{\theta}})\) and the associated changepoint configuration \(\hat{C}\) for the best-fitting model can also be extracted, as illustrated below.

\begin{verbatim}
res_cptga <- cptga(
  ObjFunc = arima_bic, N = N, popSize = 100, pcrossover = 0.95,
  pmutation = 0.3, pchangepoint = 10 / N, seed = 1234,
  XMat = XMatT, Xt = Xt
)
summary(res_cptga)
\end{verbatim}

\begin{verbatim}
#> ###############################################
#> #         Changepoint Detection via GA        #
#> ###############################################
#>    Settings: 
#>    Population size         =  100 
#>    Number of generations   =  5271 
#>    Crossover probability   =  0.95 
#>    Mutation probability    =  0.3 
#>    Changepoint probability =  0.01 
#>    minDist                 =  1 
#>    Task mode               =  cp 
#>    Parallel Usage          =  FALSE 
#>    Seed                    =  1234 
#> 
#> ##### GA results ##### 
#>    Optimal Fitness value = 68.54954 
#>    Optimal Solution: 
#>         Number of Changepoints =  2 
#>         Changepoints Locations =  250 750
\end{verbatim}

\begin{verbatim}
fit_cptga <- res_cptga@overbestfit
fit_cptga
\end{verbatim}

\begin{verbatim}
#> [1] 68.54954
\end{verbatim}

\begin{verbatim}
tau_cptga <- res_cptga@overbestchrom
tau_cptga
\end{verbatim}

\begin{verbatim}
#> [1]    2  250  750 1001
\end{verbatim}

\noindent In this example, the basic genetic algorithm detected 2 changepoints, \(\hat{\boldsymbol{\tau}} = \{\text{250, 750}\}\), with the chromosome ending flag of 1001. The minimized BIC is 68.55.

For the island model genetic algorithm, we maintain an initial population size of 100 and set \texttt{numIslands\ =\ 5}, yielding 20 individuals per island. To align the stopping criterion with \texttt{cptga} -- where the algorithm terminates once the overall best-fitting value remains unchanged for 5000 consecutive generations -- we specify \texttt{maxgen\ =\ 50} and \texttt{maxconv\ =\ 100} for \texttt{cptgaisl}. The same genetic operators are applied to generate new individuals at each generation. The estimated changepoint configuration and fitness value can be extracted in the same manner.

\begin{verbatim}
res_cptgaisl <- cptgaisl(
  ObjFunc = arima_bic, N = N, popSize = 100, numIslands = 5,
  pcrossover = 0.95, pmutation = 0.3, pchangepoint = 10 / N,
  seed = 1234, XMat = XMatT, Xt = Xt
)
summary(res_cptgaisl)
\end{verbatim}

\begin{verbatim}
#> ###############################################
#> #  Changepoint Detection via Island Model GA  #
#> ###############################################
#>    Settings: 
#>    Population size         =  100 
#>    Number of Island        =  5 
#>    Island size             =  20 
#>    Number of generations   =  5150 
#>    Number of migrations    =  103 
#>    Crossover probability   =  0.95 
#>    Mutation probability    =  0.3 
#>    Changepoint probability =  0.01 
#>    minDist                 =  1 
#>    Task mode               =  cp 
#>    Parallel Usage          =  FALSE 
#>    Seed                    =  1234 
#> 
#> ##### Island Model GA results ##### 
#>    Optimal Fitness value = 68.54954 
#>    Optimal Solution: 
#>         Number of Changepoints =  2 
#>         Changepoints Locations =  250 750
\end{verbatim}

\begin{verbatim}
fit_cptgaisl <- res_cptgaisl@overbestfit
fit_cptgaisl
\end{verbatim}

\begin{verbatim}
#> [1] 68.54954
\end{verbatim}

\begin{verbatim}
tau_cptgaisl <- res_cptgaisl@overbestchrom
tau_cptgaisl
\end{verbatim}

\begin{verbatim}
#> [1]    2  250  750 1001
\end{verbatim}

\noindent The island model genetic algorithm detected 2 changepoints, \(\hat{\boldsymbol{\tau}} = \{\text{250, 750}\}\), with the chromosome ending flag of 1001. The minimized BIC is 68.55, identical to that obtained with the basic genetic algorithm in this run. As genetic algorithms are stochastic search methods, results may vary across executions.

By comparing the changepoints estimated by the basic genetic algorithm and the island model genetic algorithm with the true values \((250, 750)\) used in the simulation, discrepancies in both number and locations can be quantified using the distance metric implemented in \texttt{cpt\_dist}.

\begin{verbatim}
m_cptga <- tau_cptga[1]
tau_cptga <- tau_cptga[2:(1 + m_cptga)]
dist_cptga <- cpt_dist(tau1 = tau_cptga, tau2 = c(250, 750), N = N)
dist_cptga
\end{verbatim}

\begin{verbatim}
#> [1] 0
\end{verbatim}

\begin{verbatim}
m_cptgaisl <- tau_cptgaisl[1]
tau_cptgaisl <- tau_cptgaisl[2:(1 + m_cptgaisl)]
dist_cptgaisl <- cpt_dist(tau1 = tau_cptgaisl, tau2 = c(250, 750), N = N)
dist_cptgaisl
\end{verbatim}

\begin{verbatim}
#> [1] 0
\end{verbatim}

\noindent Both \texttt{cptga} and \texttt{cptgaisl} produced changepoint configurations closely matching the ground truth. No substantial discrepancies were observed between the estimated and true changepoint locations used in data generation, yielding \texttt{cpt\_dist} values near zero.

We next compare the robustness and computational efficiency of \CRANpkg{changepointGA} and \CRANpkg{GA}. Using the same simulation setup as above, we generated 1000 AR(1) time series and applied \texttt{cptga} and \texttt{cptgaisl} from \CRANpkg{changepointGA}, as well as \texttt{ga} and \texttt{gaisl} from \CRANpkg{GA}, to each simulated series. Here, \texttt{cptga} and \texttt{ga} implement the basic genetic algorithm, while \texttt{cptgaisl} and \texttt{gaisl} implement the island model genetic algorithm.

Five scenarios were considered, differing by package, population size, and stopping criterion, as summarized in Table \ref{tab:ScenarioTable}. Each scenario is labeled in the format Package--Population size--Stopping criteria, where ``GA'' denotes the \CRANpkg{GA} package and ``CG'' denotes the \CRANpkg{changepointGA} package. Population sizes are categorized as ``S'' (small, 100 individuals), ``M'' (medium, 400 individuals), and ``L'' (large, 800 individuals), while stopping criteria are indicated as ``1K'' (1000) or ``5K'' (5000), representing the maximum number of consecutive generations or migrations without fitness improvement before termination. For example, CG-S-1K corresponds to \CRANpkg{changepointGA} with a small population of 100 and a stopping criterion of 1000.

\begin{table}[!h]
\centering
\caption{\label{tab:unnamed-chunk-5}Different hyperparameter settings for the basic and island model genetic algorithm implementations in the GA and changepointGA packages.\label{tab:ScenarioTable}}
\centering
\begin{tabular}[t]{lrrl}
\toprule
Scenarios & Population size & Stopping criteria & Package\\
\midrule
\cellcolor{gray!10}{GA-S-1K} & \cellcolor{gray!10}{100} & \cellcolor{gray!10}{1000} & \cellcolor{gray!10}{GA}\\
CG-S-1K & 100 & 1000 & changepointGA\\
\cellcolor{gray!10}{CG-S-5K} & \cellcolor{gray!10}{100} & \cellcolor{gray!10}{5000} & \cellcolor{gray!10}{changepointGA}\\
CG-M-5K & 400 & 5000 & changepointGA\\
\cellcolor{gray!10}{CG-L-5K} & \cellcolor{gray!10}{800} & \cellcolor{gray!10}{5000} & \cellcolor{gray!10}{changepointGA}\\
\bottomrule
\end{tabular}
\end{table}

To ensure comparability, key hyperparameters were aligned across the basic and island model genetic algorithm runs in both packages. In the island model genetic algorithm, populations were evenly distributed across islands: 100 individuals across five islands, 400 across 20 islands, and 800 across 40 islands, maintaining a fixed island size of 20. Migration occurred every 50 generations. For scenarios with a 1K stopping criterion, convergence was assessed after 20 consecutive migrations without fitness improvement; for 5K, the maximum number of consecutive migrations was increased to 100. As the scenario progressed from CG-S-1K to CG-L-5K, both population size and the stopping criterion increased. These more intensive settings were expected to yield solutions closer to the global optimum by providing a larger search budget and improved convergence behavior.

The fitness function \texttt{ARIMA\_BIC\_GA}, provided in the supplement materials, closely mirrors \texttt{arima\_bic} from \CRANpkg{changepointGA}, with the primary distinction being the initial chromosome structure, which was adapted to match the input format required by \CRANpkg{GA}. In addition, three values of the minimum changepoint spacing parameter were examined, with \texttt{minDist}=1 used as the default in both packages. Since our \CRANpkg{changepointGA} package allows variation in minimum segment length -- a unique feature of \CRANpkg{changepointGA} -- larger values (\texttt{minDist} = 3, 5) were also tested to investigate the impact of imposing a minimum segment length on changepoint detection.

\begin{figure}[h!]

{\centering \includegraphics[width=0.8\linewidth]{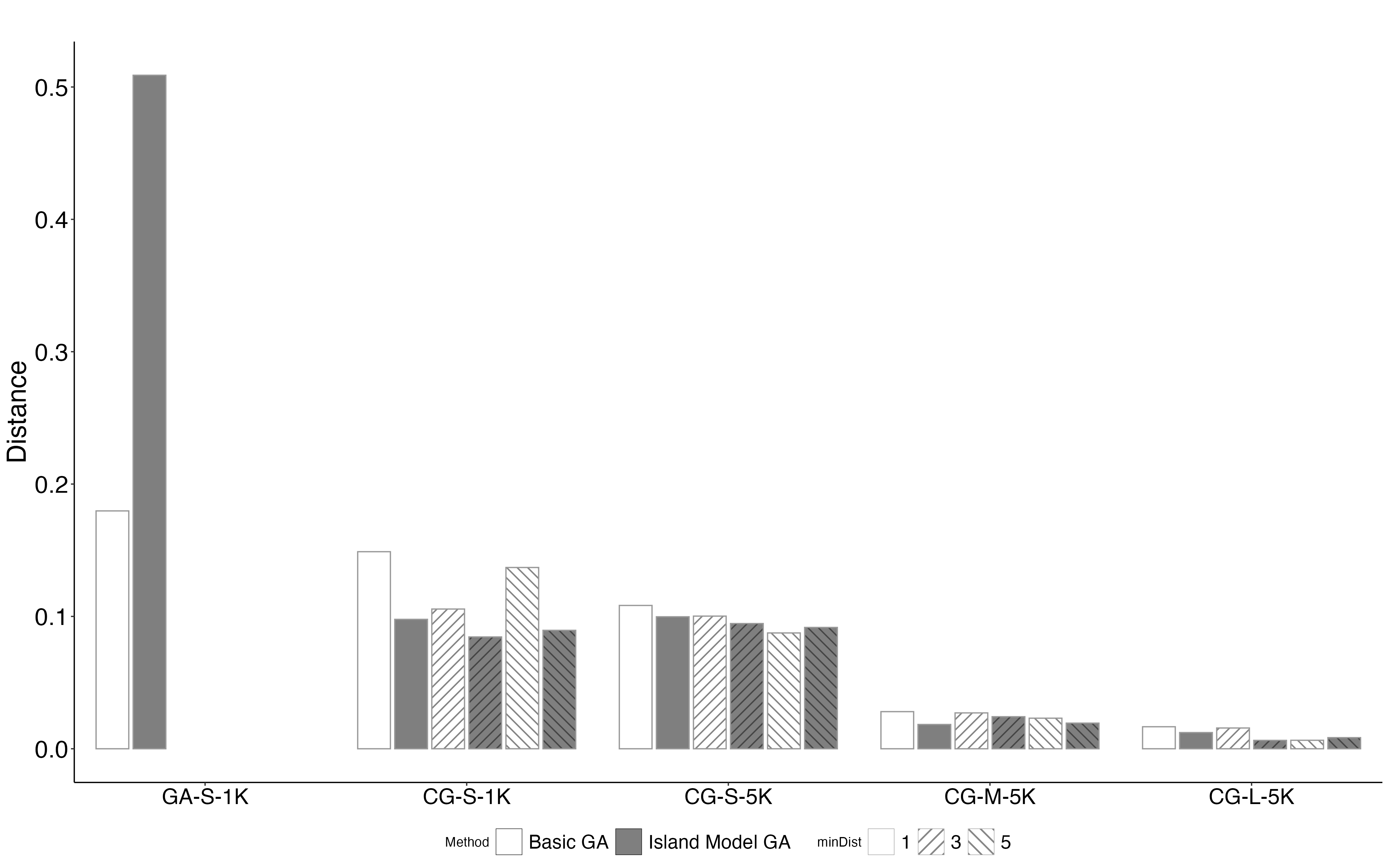} 

}

\caption{Barplots showing the average distance between true and estimated changepoint locations for the basic genetic algorithm (Basic GA) and the island model genetic algorithm (Island Model GA) under different simulation scenarios.}\label{fig:pkgcomparefig}
\end{figure}

The performance of changepoint detection was evaluated in terms of detection accuracy, computational efficiency, and optimized fitness function values. Figure \ref{fig:pkgcomparefig} illustrates detection accuracy by plotting the average distance, as evaluated by the \code{cpt\_dist} function, between the estimated changepoint configuration and the true configuration \(C = (2, 250, 750, 1001)'\) across 1000 simulations. For each of the four scenarios using \CRANpkg{changepointGA}, results are reported for three minimum changepoint spacing values (\texttt{minDist} = 1, 3, 5), whereas for the GA-S-1K scenario only \texttt{minDist} = 1 is reported (since this option is not available in \CRANpkg{GA}). Across the five scenarios, both the basic and island model genetic algorithm implementations in \CRANpkg{changepointGA} achieved smaller average distances than GA-S-1K in \CRANpkg{GA}, indicating higher detection accuracy. In general, as \texttt{minDist} increases, \CRANpkg{changepointGA} produced slightly smaller average distances, with the best performance observed under the most intensive search setting (CG-L-5K) with \texttt{minDist\ =\ 5}.

Table \ref{tab:CompTimeTab} summarizes computational efficiency by comparing average execution times. By narrowing the search space through the chromosome design in Equation \eqref{eq:chromosome}, the \CRANpkg{changepointGA} implementations substantially reduced computation time relative to \CRANpkg{GA}. For instance, with \texttt{minDist} = 1, the average runtime for the GA-S-1K scenario was approximately 3500 seconds, compared with about one second for CG-S-1K -- an almost 3500-fold reduction. Even with larger population sizes and more stringent stopping criteria, such as CG-L-5K, the average runtime remained about 535 times shorter for the basic genetic algorithm and 161 times shorter for the island model genetic algorithm relative to GA-S-1K. For \texttt{minDist} = 3 and 5, the search space was further reduced by excluding unrealistic changepoint configurations, dropping the CG-S-1K average runtime to 0.30 seconds for the basic genetic algorithm and 0.70 seconds for the island model genetic algorithm.

Across \CRANpkg{changepointGA} scenarios, the island model genetic algorithm consistently improved detection performance relative to the basic genetic algorithm (Figure \ref{fig:pkgcomparefig}), with only a small runtime increase (Table \ref{tab:CompTimeTab}). As expected, increasing population size and tightening the stopping criterion raised execution time but decreased average distance, thereby improving detection accuracy.

\begin{table}[!h]
\centering
\caption{\label{tab:unnamed-chunk-6}Average computation time (seconds) from 1000 simulations of basic genetic algorithm (Basic GA) and island model genetic algorithm (Island model GA) across five scenarios and three minimum changepoint spacing values.\label{tab:CompTimeTab}}
\centering
\fontsize{6}{8}\selectfont
\begin{tabular}[t]{>{}l|rrrr>{}r|rrrrr}
\toprule
\multicolumn{1}{c}{ } & \multicolumn{5}{c}{Basic GA} & \multicolumn{5}{c}{Island model GA} \\
\cmidrule(l{3pt}r{3pt}){2-6} \cmidrule(l{3pt}r{3pt}){7-11}
minDist & GA-S-1K & CG-S-1K & CG-S-5K & CG-M-5K & CG-L-5K & GA-S-1K & CG-S-1K & CG-S-5K & CG-M-5K & CG-L-5K\\
\midrule
\cellcolor{gray!10}{1} & \cellcolor{gray!10}{3512.69} & \cellcolor{gray!10}{0.98} & \cellcolor{gray!10}{1.92} & \cellcolor{gray!10}{3.78} & \cellcolor{gray!10}{6.56} & \cellcolor{gray!10}{3428.14} & \cellcolor{gray!10}{1.25} & \cellcolor{gray!10}{5.07} & \cellcolor{gray!10}{19.98} & \cellcolor{gray!10}{21.23}\\
3 & --- & 0.30 & 0.83 & 1.62 & 2.62 & --- & 0.70 & 2.85 & 11.36 & 11.92\\
\cellcolor{gray!10}{5} & \cellcolor{gray!10}{---} & \cellcolor{gray!10}{0.30} & \cellcolor{gray!10}{0.84} & \cellcolor{gray!10}{1.58} & \cellcolor{gray!10}{2.63} & \cellcolor{gray!10}{---} & \cellcolor{gray!10}{0.70} & \cellcolor{gray!10}{2.90} & \cellcolor{gray!10}{11.51} & \cellcolor{gray!10}{12.08}\\
\bottomrule
\end{tabular}
\end{table}

To further evaluate changepoint detection performance, the empirical frequency distributions of the number of estimated changepoints are reported in Table \ref{tab:locationestTab}. For both the basic and island model genetic algorithm implementations in \CRANpkg{changepointGA}, the true value of two changepoints was selected more frequently as the search budget increases. The percentage of simulations (out of 1000) yielding the correct number of changepoints with \CRANpkg{changepointGA} was generally higher than that for the GA-S-1K scenario, except for CG-S-1K with \code{minDist = 1}.  Although GA-S-1K appears more concentrated around $m = 2$ than CG-S-1K for the basic genetic algorithm, the overall distribution for GA-S-1K is more dispersed and yields a larger average changepoint-location distance, as shown in Figure \ref{fig:pkgcomparefig}. Under a small search budget, the basic genetic algorithm implementation in \CRANpkg{changepointGA} occasionally selected near-miss models, such as $m = 3$, more often than GA-S-1K. This behavior may reflect the rapid convergence of the compact integer encoding to nearby configurations when the search budget is limited. The island model implementation mitigated this behavior, with \CRANpkg{changepointGA} selecting the true value $m = 2$ more consistently.

Moreover, the results in Table~\ref{tab:locationestTab} show that the island model genetic algorithm in \CRANpkg{changepointGA} consistently improved the accuracy of estimating the number of changepoints relative to the corresponding basic genetic algorithm implementation across values of \code{minDist} and most search settings. Increasing \code{minDist} from 1 to 3 or 5 generally maintained or improved the frequency with which the true number of changepoints, \(m=2\), was selected for both the basic and island model genetic algorithms, although the improvement was not strictly monotone in all cases. The effect was most apparent for CG-S-1K under the basic genetic algorithm implementation, where the accuracy increased from \(86.5\%\) at \code{minDist}=1 to \(91.0\%\) at \code{minDist}=3, before decreasing to \(88.0\%\) at \code{minDist}=5. Thus, for this smaller-search setting, \code{minDist} produced an inverted U-shaped accuracy pattern, or equivalently, a roughly U-shaped error pattern. For larger search settings, such as CG-M-5K and CG-L-5K, the correct number of changepoints was selected with high frequency regardless of \code{minDist}, suggesting that the minimum segment length constraint has a more modest effect when the search budget is sufficiently large. Overall, these results suggest that \code{minDist} is most useful as a regularization device when the search budget is limited, helping to reduce the tendency to select unrealistic nearby changepoint configurations. In real applications, we recommend choosing \code{minDist} based on prior knowledge of the shortest meaningful regime length. When such information is unavailable, a sensitivity analysis over several small values may be useful.

\begin{table}[!h]
\centering
\caption{\label{tab:unnamed-chunk-7}Estimated numbers of changepoints (percentages across 1000 simulations) for the basic genetic algorithm (Basic GA) and the island model genetic algorithm (Island model GA) across five scenarios.\label{tab:locationestTab}}
\centering
\resizebox{\ifdim\width>\linewidth\linewidth\else\width\fi}{!}{
\fontsize{6}{8}\selectfont
\begin{tabular}[t]{>{}l|>{}l|rrrr>{}r|rrrrr}
\toprule
\multicolumn{1}{c}{ } & \multicolumn{1}{c}{ } & \multicolumn{5}{c}{Basic GA} & \multicolumn{5}{c}{Island model GA} \\
\cmidrule(l{3pt}r{3pt}){3-7} \cmidrule(l{3pt}r{3pt}){8-12}
minDist & m & GA-S-1K & CG-S-1K & CG-S-5K & CG-M-5K & CG-L-5K & GA-S-1K & CG-S-1K & CG-S-5K & CG-M-5K & CG-L-5K\\
\midrule
\cellcolor{gray!10}{} & \cellcolor{gray!10}{0} & \cellcolor{gray!10}{0.0} & \cellcolor{gray!10}{0.0} & \cellcolor{gray!10}{0.0} & \cellcolor{gray!10}{0.0} & \cellcolor{gray!10}{0.0} & \cellcolor{gray!10}{0.0} & \cellcolor{gray!10}{0.0} & \cellcolor{gray!10}{0.0} & \cellcolor{gray!10}{0.0} & \cellcolor{gray!10}{0.0}\\
 & 1 & 0.0 & 0.0 & 0.3 & 0.0 & 0.0 & 0.0 & 0.0 & 0.1 & 0.0 & 0.0\\
\cellcolor{gray!10}{} & \cellcolor{gray!10}{2} & \cellcolor{gray!10}{89.8} & \cellcolor{gray!10}{86.5} & \cellcolor{gray!10}{90.1} & \cellcolor{gray!10}{97.6} & \cellcolor{gray!10}{98.7} & \cellcolor{gray!10}{75.6} & \cellcolor{gray!10}{91.4} & \cellcolor{gray!10}{91.3} & \cellcolor{gray!10}{98.6} & \cellcolor{gray!10}{99.1}\\
 & 3 & 4.7 & 12.9 & 9.7 & 2.4 & 1.3 & 3.9 & 8.2 & 8.2 & 1.4 & 0.9\\
\cellcolor{gray!10}{1} & \cellcolor{gray!10}{4} & \cellcolor{gray!10}{4.3} & \cellcolor{gray!10}{0.6} & \cellcolor{gray!10}{0.2} & \cellcolor{gray!10}{0.0} & \cellcolor{gray!10}{0.0} & \cellcolor{gray!10}{17.2} & \cellcolor{gray!10}{0.4} & \cellcolor{gray!10}{0.5} & \cellcolor{gray!10}{0.0} & \cellcolor{gray!10}{0.0}\\
\addlinespace
 & 5 & 0.6 & 0.0 & 0.0 & 0.0 & 0.0 & 1.5 & 0.0 & 0.0 & 0.0 & 0.0\\
\cellcolor{gray!10}{} & \cellcolor{gray!10}{6} & \cellcolor{gray!10}{0.5} & \cellcolor{gray!10}{0.0} & \cellcolor{gray!10}{0.0} & \cellcolor{gray!10}{0.0} & \cellcolor{gray!10}{0.0} & \cellcolor{gray!10}{1.6} & \cellcolor{gray!10}{0.0} & \cellcolor{gray!10}{0.0} & \cellcolor{gray!10}{0.0} & \cellcolor{gray!10}{0.0}\\
 & 7 & 0.1 & 0.0 & 0.0 & 0.0 & 0.0 & 0.1 & 0.0 & 0.0 & 0.0 & 0.0\\
\cellcolor{gray!10}{} & \cellcolor{gray!10}{>7} & \cellcolor{gray!10}{0.0} & \cellcolor{gray!10}{0.0} & \cellcolor{gray!10}{0.0} & \cellcolor{gray!10}{0.0} & \cellcolor{gray!10}{0.0} & \cellcolor{gray!10}{0.1} & \cellcolor{gray!10}{0.0} & \cellcolor{gray!10}{0.0} & \cellcolor{gray!10}{0.0} & \cellcolor{gray!10}{0.0}\\
\midrule
 & 0 & --- & 0.0 & 0.0 & 0.0 & 0.0 & --- & 0.0 & 0.0 & 0.0 & \vphantom{1} 0.0\\
\addlinespace
\cellcolor{gray!10}{} & \cellcolor{gray!10}{1} & \cellcolor{gray!10}{---} & \cellcolor{gray!10}{0.0} & \cellcolor{gray!10}{0.0} & \cellcolor{gray!10}{0.0} & \cellcolor{gray!10}{0.0} & \cellcolor{gray!10}{---} & \cellcolor{gray!10}{0.0} & \cellcolor{gray!10}{0.0} & \cellcolor{gray!10}{0.0} & \cellcolor{gray!10}{\vphantom{1} 0.0}\\
 & 2 & --- & 91.0 & 91.2 & 97.7 & 98.8 & --- & 92.4 & 91.6 & 98.0 & 99.7\\
\cellcolor{gray!10}{} & \cellcolor{gray!10}{3} & \cellcolor{gray!10}{---} & \cellcolor{gray!10}{8.2} & \cellcolor{gray!10}{8.3} & \cellcolor{gray!10}{2.3} & \cellcolor{gray!10}{1.2} & \cellcolor{gray!10}{---} & \cellcolor{gray!10}{7.5} & \cellcolor{gray!10}{8.1} & \cellcolor{gray!10}{2.0} & \cellcolor{gray!10}{0.3}\\
3 & 4 & --- & 0.8 & 0.5 & 0.0 & 0.0 & --- & 0.1 & 0.3 & 0.0 & 0.0\\
\cellcolor{gray!10}{} & \cellcolor{gray!10}{5} & \cellcolor{gray!10}{---} & \cellcolor{gray!10}{0.0} & \cellcolor{gray!10}{0.0} & \cellcolor{gray!10}{0.0} & \cellcolor{gray!10}{0.0} & \cellcolor{gray!10}{---} & \cellcolor{gray!10}{0.0} & \cellcolor{gray!10}{0.0} & \cellcolor{gray!10}{0.0} & \cellcolor{gray!10}{\vphantom{1} 0.0}\\
\addlinespace
 & 6 & --- & 0.0 & 0.0 & 0.0 & 0.0 & --- & 0.0 & 0.0 & 0.0 & \vphantom{1} 0.0\\
\cellcolor{gray!10}{} & \cellcolor{gray!10}{7} & \cellcolor{gray!10}{---} & \cellcolor{gray!10}{0.0} & \cellcolor{gray!10}{0.0} & \cellcolor{gray!10}{0.0} & \cellcolor{gray!10}{0.0} & \cellcolor{gray!10}{---} & \cellcolor{gray!10}{0.0} & \cellcolor{gray!10}{0.0} & \cellcolor{gray!10}{0.0} & \cellcolor{gray!10}{\vphantom{1} 0.0}\\
 & >7 & --- & 0.0 & 0.0 & 0.0 & 0.0 & --- & 0.0 & 0.0 & 0.0 & \vphantom{1} 0.0\\
\midrule
\cellcolor{gray!10}{} & \cellcolor{gray!10}{0} & \cellcolor{gray!10}{---} & \cellcolor{gray!10}{0.0} & \cellcolor{gray!10}{0.0} & \cellcolor{gray!10}{0.0} & \cellcolor{gray!10}{0.0} & \cellcolor{gray!10}{---} & \cellcolor{gray!10}{0.0} & \cellcolor{gray!10}{0.0} & \cellcolor{gray!10}{0.0} & \cellcolor{gray!10}{0.0}\\
 & 1 & --- & 0.0 & 0.0 & 0.0 & 0.0 & --- & 0.0 & 0.0 & 0.0 & 0.0\\
\addlinespace
\cellcolor{gray!10}{} & \cellcolor{gray!10}{2} & \cellcolor{gray!10}{---} & \cellcolor{gray!10}{88.0} & \cellcolor{gray!10}{92.2} & \cellcolor{gray!10}{98.1} & \cellcolor{gray!10}{99.7} & \cellcolor{gray!10}{---} & \cellcolor{gray!10}{92.3} & \cellcolor{gray!10}{92.0} & \cellcolor{gray!10}{98.5} & \cellcolor{gray!10}{99.5}\\
 & 3 & --- & 11.1 & 7.6 & 1.9 & 0.3 & --- & 7.2 & 7.6 & 1.5 & 0.5\\
\cellcolor{gray!10}{5} & \cellcolor{gray!10}{4} & \cellcolor{gray!10}{---} & \cellcolor{gray!10}{0.9} & \cellcolor{gray!10}{0.2} & \cellcolor{gray!10}{0.0} & \cellcolor{gray!10}{0.0} & \cellcolor{gray!10}{---} & \cellcolor{gray!10}{0.5} & \cellcolor{gray!10}{0.4} & \cellcolor{gray!10}{0.0} & \cellcolor{gray!10}{0.0}\\
 & 5 & --- & 0.0 & 0.0 & 0.0 & 0.0 & --- & 0.0 & 0.0 & 0.0 & 0.0\\
\cellcolor{gray!10}{} & \cellcolor{gray!10}{6} & \cellcolor{gray!10}{---} & \cellcolor{gray!10}{0.0} & \cellcolor{gray!10}{0.0} & \cellcolor{gray!10}{0.0} & \cellcolor{gray!10}{0.0} & \cellcolor{gray!10}{---} & \cellcolor{gray!10}{0.0} & \cellcolor{gray!10}{0.0} & \cellcolor{gray!10}{0.0} & \cellcolor{gray!10}{0.0}\\
\addlinespace
 & 7 & --- & 0.0 & 0.0 & 0.0 & 0.0 & --- & 0.0 & 0.0 & 0.0 & 0.0\\
\cellcolor{gray!10}{} & \cellcolor{gray!10}{>7} & \cellcolor{gray!10}{---} & \cellcolor{gray!10}{0.0} & \cellcolor{gray!10}{0.0} & \cellcolor{gray!10}{0.0} & \cellcolor{gray!10}{0.0} & \cellcolor{gray!10}{---} & \cellcolor{gray!10}{0.0} & \cellcolor{gray!10}{0.0} & \cellcolor{gray!10}{0.0} & \cellcolor{gray!10}{0.0}\\
\bottomrule
\end{tabular}}
\end{table}

\begin{figure}[h!]

{\centering \includegraphics[width=0.8\linewidth]{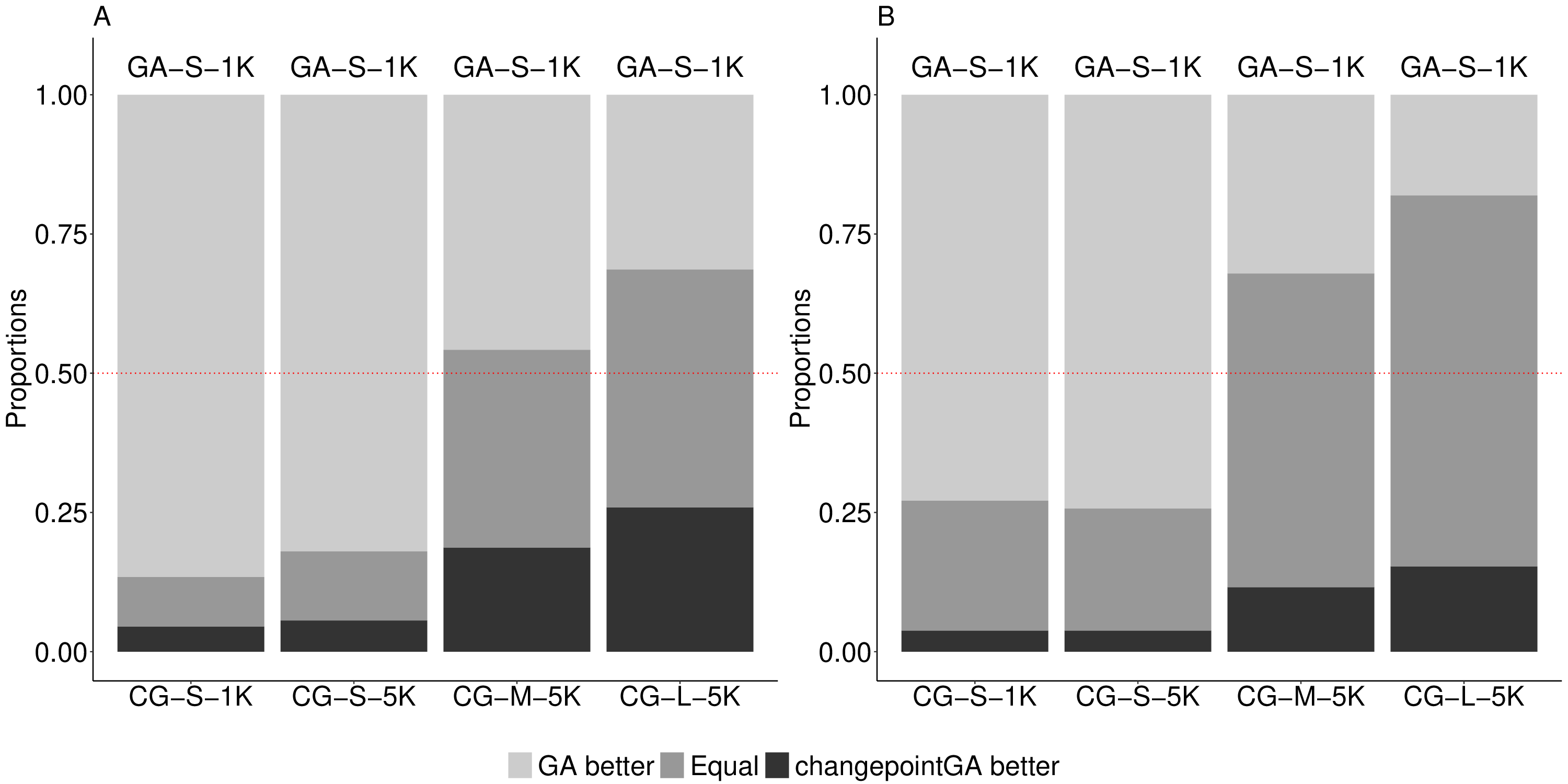} 

}

\caption{Stacked barplots comparing optimized BIC values between the GA-S-1K scenario from the GA package and the CG-S-1K, CG-S-5K, CG-M-5K, and CG-L-5K scenarios from changepointGA, using the basic genetic algorithm (A) and island model genetic algorithm (B).}\label{fig:optimcomparefig}
\end{figure}

Furthermore, we investigate whether \CRANpkg{changepointGA} or \CRANpkg{GA} attains the minimum BIC value more frequently under the fair comparison setting with \texttt{minDist\ =\ 1}. Figure \ref{fig:optimcomparefig} presents stacked barplots comparing the optimized fitness function values between GA-S-1K and each scenario using \CRANpkg{changepointGA} for both the basic genetic algorithm in panel A and the island model genetic algorithm in panel B. For each simulation run, there are three possible outcomes: if \CRANpkg{changepointGA} achieves a smaller BIC, the run is categorized as ``changepointGA better''; if both \CRANpkg{changepointGA} and \CRANpkg{GA} yield identical BICs, the outcome is labeled ``Equal''; and if \CRANpkg{GA} yields a smaller BIC, the run is categorized as ``GA better''.

It is evident from Figure \ref{fig:optimcomparefig} that \CRANpkg{changepointGA} sacrifices a degree of optimization accuracy in exchange for faster runtime. For example, CG-L-5K and GA-S-1K agree most frequently, with CG-L-5K attaining the lowest BIC more often in both the basic genetic algorithm and island model genetic algorithm implementations among scenarios using \CRANpkg{changepointGA}. Although the population size increases from 100 (GA-S-1K) to 800 (CG-L-5K) with more stringent stopping criteria, \CRANpkg{changepointGA} achieves comparable optimization performance, with a much shorter average distance and a much higher rate of correctly detecting two changepoints (true \(m = 2\)). Most importantly, it runs over 150 times faster.

\subsubsection{4.1.2 Six changepoints}\label{six-changepoints}

To further assess robustness, we conducted an additional simulation study under a more challenging multiple changepoint setting. The ARMA orders were treated as known, with \(p=1\) and \(q=0\). In all cases, the series length was fixed at \(T=1000\), and six true changepoint locations were set at \(200, 400, 500, 700, 750,\) and \(850\). The corresponding mean shift magnitudes were specified as $a \times (1,2,1,0,1,2)$, where \(a \in \{1,2,3\}\) controls the overall signal strength. The error process was generated from the same AR(1) model used above, with \(\phi=0.5\) and \(\sigma^2=1\).

\begin{figure}[h!]

{\centering \includegraphics[width=0.8\linewidth]{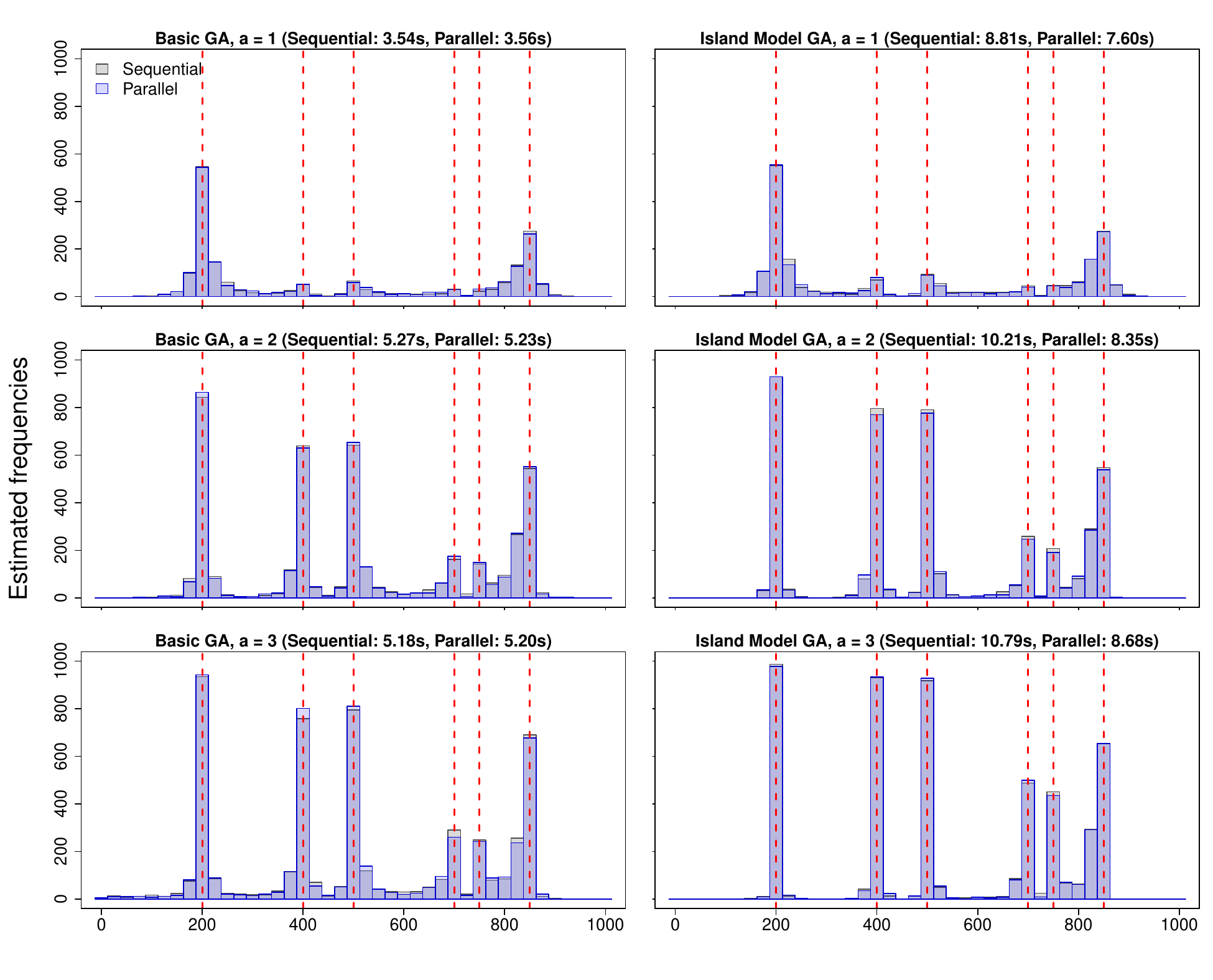} 

}

\caption{Histograms of the estimated changepoint locations with known ARMA orders, using the basic genetic algorithm (Basic GA in the left column) and the island model genetic algorithm (Island Model GA in the right column). Red dashed lines indicate the true changepoint positions; gray and blue histograms show sequential and parallel results, respectively.}\label{fig:6chpts}
\end{figure}

For each signal setting, 1000 simulated series were generated. Both \code{cptga} and \code{cptgaisl} were run using sequential computation (\code{parallel = FALSE}) and parallel computation (\code{parallel = TRUE} with \code{nCore = 10}). This design allows us to assess how detection accuracy varies with signal strength, with particular attention to whether the nearby changepoints at 700 and 750 are successfully identified, and how much runtime can be reduced through parallelization. The estimated changepoint locations from the basic genetic algorithm and island model genetic algorithm implementations are displayed in Figure \ref{fig:6chpts}.

The histograms in Figure~\ref{fig:6chpts} show that changepoint detection improves as the signal strength \(a\) increases. For \(a=1\), both methods most clearly identify the larger shifts, whereas the smaller and nearby shifts, especially those around 700 and 750, are detected less consistently. For \(a=2\) and \(a=3\), the estimated changepoint locations become more concentrated around the true changepoints, with the island model genetic algorithm generally producing sharper peaks than the basic genetic algorithm. The sequential and parallel implementations yield nearly identical detection patterns, indicating that parallelization primarily affects runtime rather than estimation accuracy. The main benefit of parallelization is computational: the reported runtimes show that parallel computation reduces runtime for the island model genetic algorithm, whereas the runtime savings are small for the basic genetic algorithm. Because parallelization is implemented primarily in fitness evaluations, its computational advantage is expected to be greater when the population size is large or when the objective function is expensive to evaluate.

\subsubsection{4.1.3 No changepoints}\label{no-changepoints}

Lastly, to assess false positive behavior, we considered a no-changepoint setting with series length \(T=1000\). The data were generated from a stationary AR(1) process with \(\phi=0.5\) and \(\sigma^2=1\). No mean or trend shifts were introduced, so the true number of changepoints is \(m=0\). This setting provides a baseline for evaluating whether the \CRANpkg{changepointGA} spuriously detects changepoints in a stationary time series. Among 1000 simulated series, the basic genetic algorithm correctly detected no changepoints in \(99.7\%\) of cases under both sequential and parallel computation. The island model genetic algorithm correctly detected no changepoints in \(99.3\%\) and \(99.7\%\) of cases under sequential and parallel computation, respectively.

\subsection{\texorpdfstring{4.2 ARMA(\(p\), \(q\)) orders unknown}{4.2 ARMA(p, q) orders unknown}}\label{armap-q-orders-unknown}

In practical applications, the orders of an ARMA(\(p\), \(q\)) model are typically unknown. A more robust approach is to jointly estimate the model orders, the number of changepoints, and their locations. Even without prespecifying \(p\) and \(q\), the BIC criterion in Equation \eqref{eq:MultipleBIC} can still be applied to simultaneously detect multiple changepoints and select appropriate model orders.

In this subsection, \({\epsilon_{t}}\) in Equation \eqref{eq:timeseriesmodel} is assumed to follow an ARMA(1,1) process:
\begin{equation*}
    \epsilon_{t} = \phi_{1}\epsilon_{t-1} + e_{t} + \theta_{1}e_{t-1},
\end{equation*}
where \(|\phi_{1}| < 1\) is the autoregressive parameter, \(|\theta_{1}| < 1\) is the moving average parameter, and \(\{e_t\}\) denotes a white noise process with mean zero and variance \(\sigma^{2}\). In the following illustration, the arguments \texttt{phi} and \texttt{theta} correspond to \(\phi_{1} = 0.5\) and \(\theta_{1} = 0.8\), respectively. Data are generated from the chromosome \(C = (2, 1, 1, 250, 750, 1001)'\), which encodes an ARMA(1,1) process with \(m = 2\) changepoints at times 250 and 750. All other parameters follow those in the previous subsection.

\begin{verbatim}
N <- 1000
XMatT <- matrix(1, nrow = N, ncol = 1)
Xt <- ts_sim(Ts=N, beta = 0.5, XMat = XMatT, sigma = 1, phi = 0.5, theta = 0.8,
  Delta = c(2, -2), CpLoc = c(250, 750), seed = 123456)
\end{verbatim}

When model order selection is required, the argument \texttt{plen} in the objective function \code{arima\_bic\_order\_pq} must be specified to inform the \CRANpkg{changepointGA} package. For an ARMA(1,1) model, setting \texttt{plen=2} indicates that two model orders are to be selected. Both the time series data \texttt{Xt} and the regression design matrix \texttt{XMat} are included as additional arguments. Inside \code{arima\_bic\_order\_pq}, ARMA model fitting is performed using the standard R function \code{arima}. The penalty term $mlog(N)$ in the BIC criterion is intended to reduce the tendency to overestimate the number of changepoints, particularly for time series in which unmodeled positive serial correlation may otherwise be captured by spurious changepoints.

\begin{verbatim}
arima_bic_order_pq(chromosome, plen = 2, XMat, Xt)
\end{verbatim}

In \texttt{cptga} and \texttt{cptgaisl} functions, the permissible range of each order parameter is defined through a list object \texttt{prange}, which contains integer ranges for the parameters. For example, in the code snippet below, \texttt{prange} specifies that both the AR and MA orders may take values from \(\{0, 1, 2, 3\}\). This list is then passed to the \texttt{cptga} and \texttt{cptgaisl}, where the number of model order parameters equals the length of \texttt{prange}. Additionally, for order selection, set \texttt{option="both"} in the call. To address the increased computational complexity of optimization, the population size is set to \code{popSize} = 800 in \texttt{cptga}, and to \code{popSize} = 800 with \texttt{numIslands\ =\ 10} in \texttt{cptgaisl}. With \code{minDist} = 3, the objective function \code{arima\_bic\_order\_pq} is then minimized using both the basic genetic algorithm and island model genetic algorithm, as illustrated below.

\begin{verbatim}
prange <- list(ar = c(0, 3), ma = c(0, 3))
res_cptga <- cptga(
  ObjFunc = arima_bic_order_pq, N = N, prange = prange,
  popSize = 800, pcrossover = 0.95, pmutation = 0.3,
  pchangepoint = 10 / N, option = "both", minDist = 3,
  XMat = XMatT, Xt = Xt, seed = 123456
)
summary(res_cptga)
\end{verbatim}

\begin{verbatim}
#> ###############################################
#> #         Changepoint Detection via GA        #
#> ###############################################
#>    Settings: 
#>    Population size         =  800 
#>    Number of generations   =  13508 
#>    Crossover probability   =  0.95 
#>    Mutation probability    =  0.3 
#>    Changepoint probability =  0.01 
#>    minDist                 =  3 
#>    Task mode               =  both 
#>    Parallel Usage          =  FALSE 
#>    Seed                    =  123456 
#> 
#> ##### GA results ##### 
#>    Optimal Fitness value = 2877.265 
#>    Optimal Solution: 
#>         Number of Changepoints =  2 
#>         Model hyperparameters:
#>              ar = 1 
#>              ma = 1 
#>         Changepoints Locations =  265 749
\end{verbatim}

\begin{verbatim}
res_cptgaisl <- cptgaisl(
  ObjFunc = arima_bic_order_pq, N = N, prange = prange,
  popSize = 800, numIslands = 10,
  pcrossover = 0.95, pmutation = 0.3,
  pchangepoint = 10 / N, option = "both", minDist = 3,
  XMat = XMatT, Xt = Xt, seed = 123456
)
summary(res_cptgaisl)
\end{verbatim}

\begin{verbatim}
#> ###############################################
#> #  Changepoint Detection via Island Model GA  #
#> ###############################################
#>    Settings: 
#>    Population size         =  800 
#>    Number of Island        =  10 
#>    Island size             =  80 
#>    Number of generations   =  5650 
#>    Number of migrations    =  113 
#>    Crossover probability   =  0.95 
#>    Mutation probability    =  0.3 
#>    Changepoint probability =  0.01 
#>    minDist                 =  3 
#>    Task mode               =  both 
#>    Parallel Usage          =  FALSE 
#>    Seed                    =  123456 
#> 
#> ##### Island Model GA results ##### 
#>    Optimal Fitness value = 2870.097 
#>    Optimal Solution: 
#>         Number of Changepoints =  2 
#>         Model hyperparameters:
#>              ar = 1 
#>              ma = 1 
#>         Changepoints Locations =  244 749
\end{verbatim}

Both the basic genetic algorithm and the island model genetic algorithm successfully estimated the AR and MA orders (\(p=1\) and \(q=1\)) and identified changepoints near their true locations. The island model genetic algorithm required fewer generations than the basic genetic algorithm, highlighting its computational advantage. Since genetic algorithms are stochastic search algorithms (as mentioned earlier), results may vary slightly across simulations. The distance metric function \texttt{cpt\_dist} can still be used to measure differences between changepoint configurations, as described in the previous subsection.

\begin{table}[!h]
\centering
\caption{\label{tab:unnamed-chunk-11}Estimated number of changepoints and ARMA($p$, $q$) orders (percentages across 1000 simulations) for the basic genetic algorithm (Basic GA) and the island model genetic algorithm (Island Model GA) in changepointGA. \label{tab:LocationestOrderTab}}
\centering
\fontsize{6}{8}\selectfont
\begin{tabular}[t]{lr>{}r|lr>{}r|lrr}
\toprule
\multicolumn{3}{c}{Number of changepoints} & \multicolumn{6}{c}{Orders of ARMA models} \\
\cmidrule(l{3pt}r{3pt}){1-3} \cmidrule(l{3pt}r{3pt}){4-9}
m & Basic GA & Island Model GA & (p, q) & Basic GA & Island Model GA & (p, q) & Basic GA & Island Model GA\\
\midrule
\cellcolor{gray!10}{0} & \cellcolor{gray!10}{0.6} & \cellcolor{gray!10}{0.7} & \cellcolor{gray!10}{(0, 0)} & \cellcolor{gray!10}{0.0} & \cellcolor{gray!10}{0.0} & \cellcolor{gray!10}{(2, 0)} & \cellcolor{gray!10}{0.0} & \cellcolor{gray!10}{0.0}\\
1 & 6.5 & 3.9 & (0, 1) & 0.0 & 0.0 & (2, 1) & 9.7 & 6.3\\
\cellcolor{gray!10}{2} & \cellcolor{gray!10}{83.5} & \cellcolor{gray!10}{91.7} & \cellcolor{gray!10}{(0, 2)} & \cellcolor{gray!10}{0.0} & \cellcolor{gray!10}{0.0} & \cellcolor{gray!10}{(2, 2)} & \cellcolor{gray!10}{13.8} & \cellcolor{gray!10}{4.5}\\
3 & 8.6 & 3.7 & (0, 3) & 0.9 & 0.4 & (2, 3) & 0.9 & 0.1\\
\cellcolor{gray!10}{4} & \cellcolor{gray!10}{0.8} & \cellcolor{gray!10}{0.0} & \cellcolor{gray!10}{(1, 0)} & \cellcolor{gray!10}{0.0} & \cellcolor{gray!10}{0.0} & \cellcolor{gray!10}{(3, 0)} & \cellcolor{gray!10}{0.0} & \cellcolor{gray!10}{0.0}\\
\addlinespace
5 & 0.0 & 0.0 & (1, 1) & 59.5 & 80.7 & (3, 1) & 2.8 & 0.0\\
\cellcolor{gray!10}{6} & \cellcolor{gray!10}{0.0} & \cellcolor{gray!10}{0.0} & \cellcolor{gray!10}{(1, 2)} & \cellcolor{gray!10}{9.2} & \cellcolor{gray!10}{7.8} & \cellcolor{gray!10}{(3, 2)} & \cellcolor{gray!10}{0.4} & \cellcolor{gray!10}{0.1}\\
>6 & 0.0 & 0.0 & (1, 3) & 2.6 & 0.1 & (3, 3) & 0.2 & 0.0\\
\bottomrule
\end{tabular}
\end{table}

\begin{figure}[h!]

{\centering \includegraphics[width=0.8\linewidth]{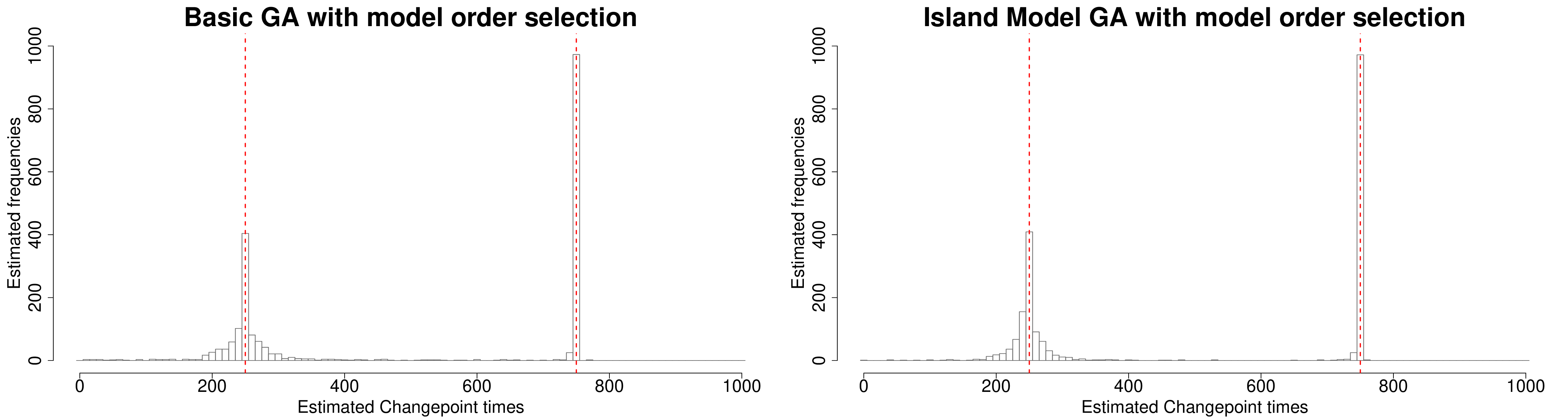} 

}

\caption{Histograms of estimated changepoint times for the basic genetic algorithm (Basic GA) and the island model genetic algorithm (Island Model GA) in changepointGA.}\label{fig:locationComparisonorder}
\end{figure}

In a simulation study of 1000 runs using the same setup as above, Table \ref{tab:LocationestOrderTab} and Figure \ref{fig:locationComparisonorder} summarize the results for both the basic genetic algorithm and the island model genetic algorithm implementations in \CRANpkg{changepointGA}.  A direct comparison with the \CRANpkg{GA} package is not available in this setting, because \CRANpkg{GA} does not provide joint estimation of ARMA orders and changepoint locations. Table \ref{tab:LocationestOrderTab} reports the empirical frequency distributions of the estimated number of changepoints and ARMA(\(p\), \(q\)) orders. The correct number of changepoints was identified in 83.5\% of runs with the basic genetic algorithm and 91.7\% with the island model genetic algorithm. Across all 16 possible ($p$, $q$) combinations from $\{0,1,2,3\}$, 59.5\% of the simulations correctly recovered the true ARMA model orders ($p = 1$, $q = 1$) for the basic genetic algorithm, and 80.7\% correctly recovered the true orders for the island model genetic algorithm.

Additionally, Figure \ref{fig:locationComparisonorder} presents histograms of the estimated changepoint locations for both the basic genetic algorithm and the island model genetic algorithm, showing the total number of times a changepoint was detected at time \(t\) for \(1 \leq t < N\). The modes of the histograms are concentrated around the true changepoint times (\(t = 250\) and \(t = 750\)). Detection at \(t = 750\) is more frequent than at \(t = 250\), reflecting the larger magnitude of change at \(t = 750\). The average distance between the estimated and true changepoint configurations was 0.185 for the basic genetic algorithm and 0.092 for the island model genetic algorithm, showing improved changepoint-location recovery under the island model. This is consistent with the island model genetic algorithm's higher recovery rates for both the correct number of changepoints and the true ARMA($1,1$) order. Overall, the \CRANpkg{changepointGA} package provides robust joint estimation of both changepoints and ARMA model orders.

\section{Applications}\label{applications}

\subsection{5.1 Glioblastoma aCGH data}\label{glioblastoma-acgh-data}

Glioblastoma multiforme (GBM) can induce chromosomal aberrations during DNA replication. Array comparative genomic hybridization (aCGH) experiments analyze these aberrations through fluorescence intensity ratios, which reflect DNA copy number variations: ratios greater than one indicate chromosomal gains, whereas ratios less than one indicate deletions. Changepoint detection provides an effective approach to gene sequence segmentation, improving the accuracy of genomic alteration detection and supporting advances in disease screening and treatment. The aCGH data used here replicate those reported in \citet{killick2014changepoint} and \citet{lai2005comparative}.

Following \citet{killick2014changepoint} and \citet{lai2005comparative}, we considered the piecewise constant mean shift model described in Equation \eqref{eq:timeseriesmodel}, with \(\epsilon_{t}\) assumed to follow an IID normal distribution and all trend parameters in Equation \eqref{eq:kappa} set to zero. The BIC was used to construct the objective function \texttt{lm\_search\_bic}, provided in the supplement materials. To avoid redundancy, only \texttt{cptgaisl} was employed to minimize this objective function for segmentation of the aCGH data.

\begin{verbatim}
load("data/arrayaCGH.rda")
Xt <- arrayaCGH[, 5]
N <- length(Xt)
XMat <- matrix(1, nrow = N, ncol = 1)
res_cptgaisl <- cptgaisl(
  ObjFunc = lm_search_bic, N = N, popSize = 400, numIslands = 10, 
  pcrossover = 0.95, pmutation = 0.3, pchangepoint = 10/N, 
  XMat = XMat, Xt = Xt, seed = 123456)
summary(res_cptgaisl)
\end{verbatim}

\begin{verbatim}
#> ###############################################
#> #  Changepoint Detection via Island Model GA  #
#> ###############################################
#>    Settings: 
#>    Population size         =  400 
#>    Number of Island        =  10 
#>    Island size             =  40 
#>    Number of generations   =  6600 
#>    Number of migrations    =  132 
#>    Crossover probability   =  0.95 
#>    Mutation probability    =  0.3 
#>    Changepoint probability =  0.05181347 
#>    minDist                 =  1 
#>    Task mode               =  cp 
#>    Parallel Usage          =  FALSE 
#>    Seed                    =  123456 
#> 
#> ##### Island Model GA results ##### 
#>    Optimal Fitness value = 391.2571 
#>    Optimal Solution: 
#>         Number of Changepoints =  6 
#>         Changepoints Locations =  81 85 89 96 123 133
\end{verbatim}

\begin{figure}[h!]

{\centering \includegraphics[width=0.7\linewidth]{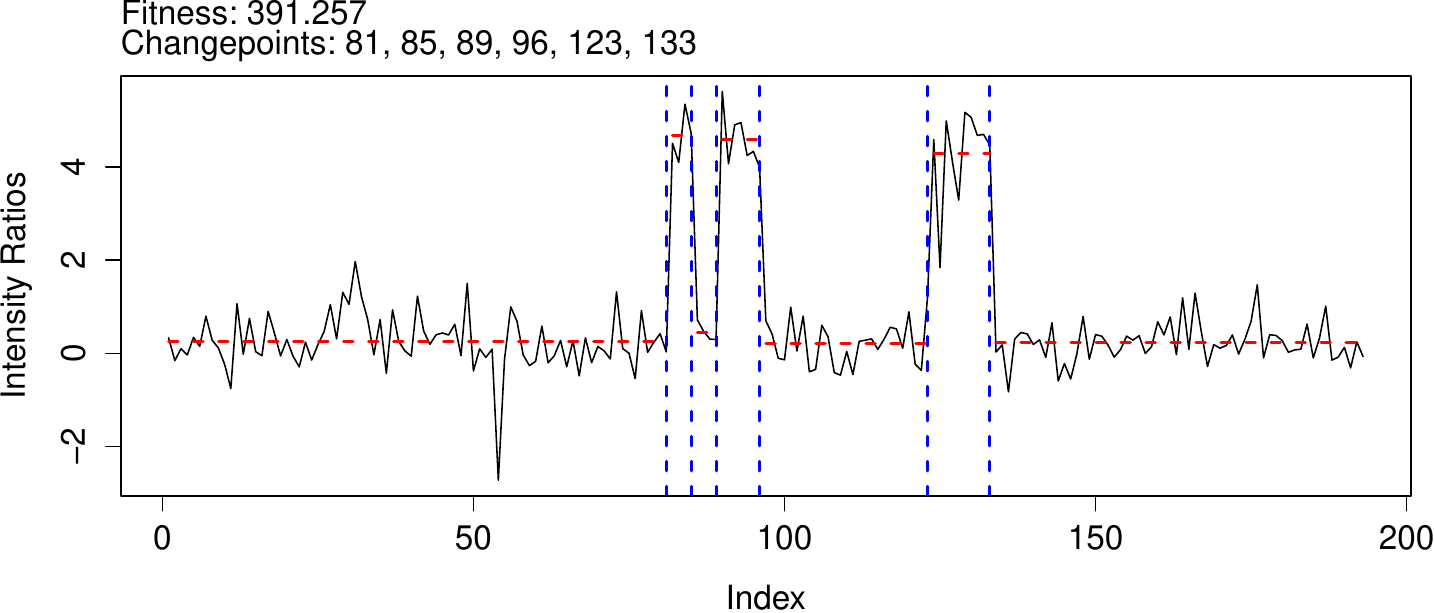} 

}

\caption{A changepoint analysis for chromosomal segments detection in Glioblastoma aCGH data.}\label{fig:CGHfig}
\end{figure}

There are \(\hat{m}=\text{6}\) detected changepoints in the sequence, indicating DNA copy number variations at \(\hat{\boldsymbol{\tau}}=\{\text{81, 85, 89, 96, 123, 133}\}\). These results closely align with those obtained from other segmentation methods, such as CGHseg \citep{killick2014changepoint} and PELT \citep{lai2005comparative}. Our implementation successfully recovered all changepoints identified by these methods and additionally detected \(\{\text{89}\}\) as a changepoint. The additional changepoint was also detected by the wavelet \citep{hsu2005denoising}, GLAD \citep{hupe2004analysis}, CLAC \citep{wang2005method}, and Quantreg \citep{eilers2005quantile} segmentation approaches implemented in \citet{killick2014changepoint}. The sequence array profile with the detected changepoints is shown in Figure \ref{fig:CGHfig}.

\begin{verbatim}
plot(res_cptgaisl, data = Xt, XAxisLab = "Index", YAxisLab = "Intensity Ratios")
\end{verbatim}

\subsection{5.2 Atlanta airport temperature series}\label{atlanta-airport-temperature-series}

In this example, we demonstrate the importance of simultaneously selecting time series model orders and detecting changepoints. The annual mean temperature data from 1879 to 2013 at Atlanta, Georgia's Hartsfield International Airport weather station, available in \citet{lund2023good}, were reanalyzed using our \CRANpkg{changepointGA} package.

\begin{verbatim}
load("data/AtlantaTemperature.rda")
Xt <- AtlantaTemperature[, 2]
N <- length(Xt)
\end{verbatim}

Three piecewise models are fitted to the Atlanta series. The first is the constant mean shift model, in which all trend parameters satisfy \(\alpha_i=0\) for \(i=0,1,\ldots,m\) in Equation~\eqref{eq:kappa}. The second is the fixed common-trend mean shift model, in which \(\alpha_i=0\) for \(i=1,\ldots,m\), so that all regimes share the baseline trend. The third is the trend shift model specified in Equation~\eqref{eq:kappa}. The error process \(\{\epsilon_t\}\) in Equation~\eqref{eq:timeseriesmodel} is assumed to follow an ARMA(\(p,q\)) process with unknown orders \(p\) and \(q\). Because \citet{lund2023good} fitted an AR(1) model to this series, we restrict the candidate AR orders to \(p \in \{0,1,2,3\}\) and set \(q=0\) to reduce computational complexity. For the constant mean shift and fixed common-trend mean shift models, the BIC objective function \code{ar\_bic\_mean} is used with different design matrices to reflect the corresponding mean structures. For the trend shift model, the BIC objective function \code{ar\_bic\_mean\_trend\_shift} is used. All three objective functions are provided in the supplementary materials. The \code{cptgaisl} method with \code{minDist=2} is then applied to each model to jointly select the AR order and detect changepoints. The results are reported in Table~\ref{tab:ModelBICTable}. Additionally, the output is displayed below only for the fixed common-trend mean shift model. 

\begin{verbatim}
prange <- list(ar = c(0, 3))

XMat_mean_fixedtrend <- cbind(1, (1:N)/N)
res_mean_fixedtrend <- cptgaisl(
  ObjFunc = ar_bic_mean, N = N, prange = prange, popSize = 400, numIslands = 10,
  minDist = 2,
  pcrossover = 0.95, pmutation = 0.3, pchangepoint = 10 / N, option = "both",
  XMat = XMat_mean_fixedtrend, Xt = Xt, seed = 123456
)
summary(res_mean_fixedtrend)
\end{verbatim}

\begin{verbatim}
#> ###############################################
#> #  Changepoint Detection via Island Model GA  #
#> ###############################################
#>    Settings: 
#>    Population size         =  400 
#>    Number of Island        =  10 
#>    Island size             =  40 
#>    Number of generations   =  5050 
#>    Number of migrations    =  101 
#>    Crossover probability   =  0.95 
#>    Mutation probability    =  0.3 
#>    Changepoint probability =  0.07462687 
#>    minDist                 =  2 
#>    Task mode               =  both 
#>    Parallel Usage          =  FALSE 
#>    Seed                    =  123456 
#> 
#> ##### Island Model GA results ##### 
#>    Optimal Fitness value = 247.2428 
#>    Optimal Solution: 
#>         Number of Changepoints =  2 
#>         Model hyperparameters:
#>              ar = 0 
#>         Changepoints Locations =  79 105
\end{verbatim}

\begin{table}[!h]
\centering
\caption{\label{tab:unnamed-chunk-20}BIC-based changepoint model selection results via the island model genetic algorithm for the Atlanta airport temperature series.\label{tab:ModelBICTable}}
\centering
\fontsize{6}{8}\selectfont
\begin{tabular}[t]{lrrr}
\toprule
Models & BIC & Detected changepoints & Detected AR order\\
\midrule
\cellcolor{gray!10}{Constant mean shift model} & \cellcolor{gray!10}{252.9865} & \cellcolor{gray!10}{1918, 1960, 1983} & \cellcolor{gray!10}{0}\\
Fixed common-trend mean shift model & 247.2428 & 1957, 1983 & 0\\
\cellcolor{gray!10}{Trend shift model} & \cellcolor{gray!10}{248.3588} & \cellcolor{gray!10}{1959} & \cellcolor{gray!10}{0}\\
\bottomrule
\end{tabular}
\end{table}

The estimated value of \(p\) is zero for all three models, indicating that the fitted error process \(\epsilon_t\) is uncorrelated. Among the three candidate models, the fixed common-trend mean shift model has the smallest BIC value, \(247.2428\). This model detects \(\hat{m}=2\) changepoints, located at
\(\hat{\boldsymbol{\tau}}=\{79,105\}\), corresponding to the years 1957 and 1983, as shown in the R output above. This illustrates that multiple changepoint detection procedures often flag one or more changepoints in an attempt to track the series mean when a common trend is present but ignored \citep{lund2023good}. By contrast, under a piecewise constant mean shift model with an AR(1) error process, \citet{lund2023good} identified three changepoints in 1921, 1960, and 1984 using the \CRANpkg{GA} package, with BIC \(=253.5776\). The \CRANpkg{changepointGA} analysis recovers two changepoints close to the later two changepoints reported by \citet{lund2023good}. The smaller BIC under the fixed common-trend specification with uncorrelated errors suggests that allowing a common trend and jointly selecting the AR order and changepoint locations provides a better fit for this series.

\begin{figure}[h!]

{\centering \includegraphics[width=0.7\linewidth]{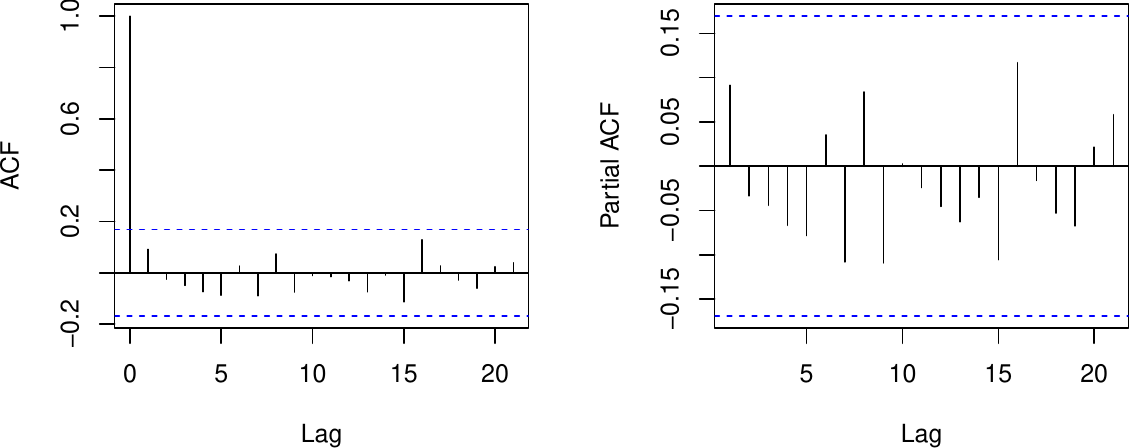} 

}

\caption{Sample ACF and PACF plots of the residuals from the fixed common-trend mean shift model with two estimated changepoints by changepointGA.}\label{fig:acfpacfAtlanta}
\end{figure}

\begin{figure}[h!]

{\centering \includegraphics[width=0.7\linewidth]{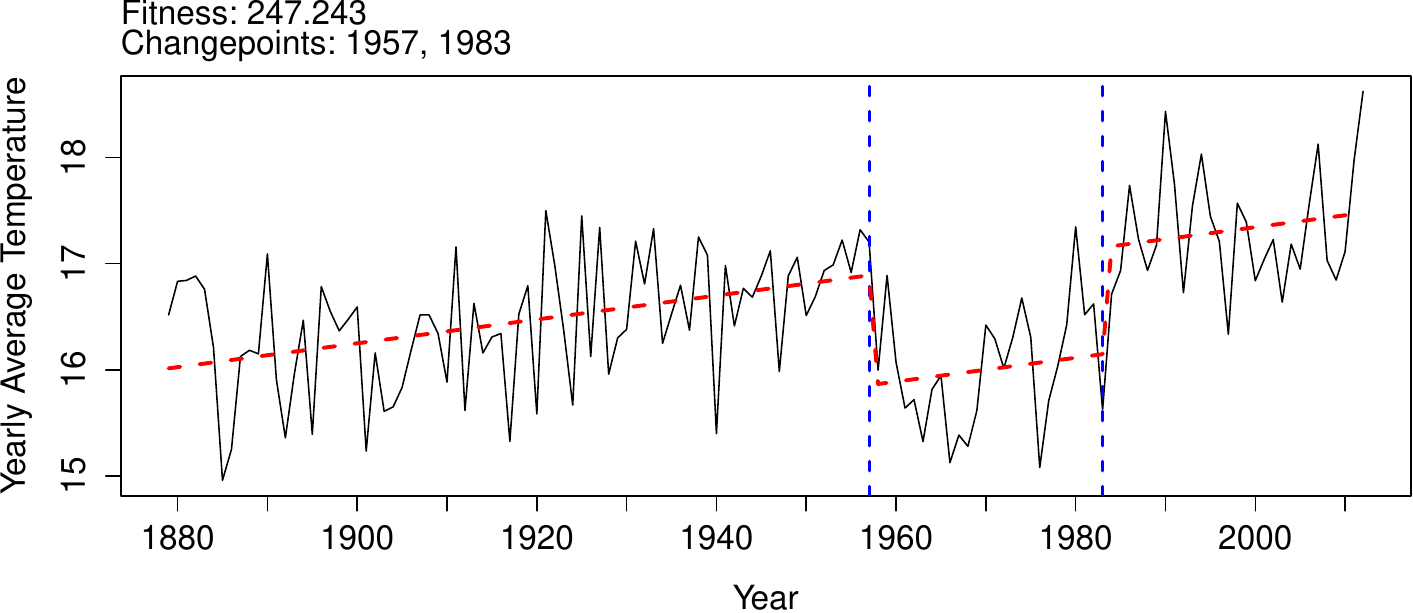} 

}

\caption{A changepoint analysis of the Atlanta airport temperature series.}\label{fig:AtlantaCpt}
\end{figure}

The sample autocorrelation function (ACF) and partial autocorrelation function (PACF) plots of the residuals from the fixed common-trend mean shift model, with two estimated changepoints at \(\hat{\boldsymbol{\tau}}=\{79,105\}\), are shown in Figure~\ref{fig:acfpacfAtlanta}. These diagnostics suggest that the fitted model adequately captures the dependence structure in the series. In addition, Figure~\ref{fig:AtlantaCpt} displays the detected changepoints, shown as blue dashed vertical lines, and the estimated level shifts around the common trend, shown as red dashed lines, overlaid on the time series. The estimated common trend parameter is \(\hat{\alpha}_0 = \text{1.5059}\), with a standard error of \(\text{0.3400}\).

\begin{verbatim}
arcptfit <- ar_bic_mean(chromosome=res_mean_fixedtrend@overbestchrom,
                        XMat=XMat_mean_fixedtrend, Xt=Xt)
DesignX <- attributes(arcptfit)$DesignX
fit <- attributes(arcptfit)$fit
beta_xreg <- tail(fit$coef, ncol(DesignX))
xbeta <- as.numeric(DesignX %*% beta_xreg)
plot(res_mean_fixedtrend, data = Xt, show_segmean = FALSE, 
     XTickLab = 1879:2012, XTickPos = seq(1880, 2010, by = 10), 
     XAxisLab = "Year", YAxisLab = "Yearly Average Temperature")
lines(1879:2012, xbeta, col = "red", lwd = 2.5, lty = "dashed")
\end{verbatim}

\section{Conclusion}\label{conclusion}

This paper presents the R package \CRANpkg{changepointGA}, which applies genetic algorithm optimization to changepoint detection problems. The package provides substantial flexibility, allowing users to customize objective functions and genetic operators, enabling experimentation and extension.

Compared with the \CRANpkg{GA} package, \CRANpkg{changepointGA} differs in three main respects. First, it employs a more compact chromosome representation for long time series, using a parameter vector whose dimensionality adapts to the chromosome configuration, whereas \CRANpkg{GA} relies on binary encoding. The choice of chromosome representation plays an important role in the performance of genetic algorithms. Second, \CRANpkg{changepointGA} enforces a minimum segment length, preventing unrealistic changepoint configurations permitted in the \CRANpkg{GA} package. Third, it enables joint estimation of time series AR and MA orders and changepoint locations, ensuring identification under the best-fitting model specification. Although \CRANpkg{changepointGA} may sacrifice some precision in achieving global optimality, it provides a computationally efficient stochastic search for changepoint detection.

We believe that the \CRANpkg{changepointGA} package will benefit the research community by providing a user-friendly, accurate, and adaptable tool for a wide range of changepoint detection applications. Its versatility makes it applicable to diverse domains, including climate studies, genetic research, quality control, and other areas requiring precise and efficient changepoint detection.




\title{Supplementary Materials for ``changepointGA: An R package for Fast Changepoint Detection via Genetic Algorithms''}

\author{by Mo Li and QiQi Lu}

\maketitle

\section{\texorpdfstring{S1. At most one changepoint (AMOC) detection using \CRANpkg{changepointGA} \label{sec:supp_amoc}}{S1. At most one changepoint (AMOC) detection using  }}\label{s1.-at-most-one-changepoint-amoc-detection-using}

Assuming a single changepoint (\(m=1\)) at time \(\tau=\{2,\ldots,N-1\}\) in the first order autoregressive (AR(1)) series \(\{X_{t}\}_{t=1}^{N}\) and the changepoint \(\tau\) partitions the process into two regimes. Such a problem often involves distinguishing between two situations: the time series contains no changepoint, or there is exactly one changepoint at an unknown time \(\tau\). Hypothesis testing is one of the prevalent methods for addressing AMOC detection problems. Specifically, the likelihood ratio test is commonly employed to discern the presence of a changepoint and ascertain its location if one indeed exists. In general, the test statistic takes the form
\begin{equation}
    \Lambda_{\text{max}}=\max_{\ell\leq\tau\leq h}\Lambda(\tau) \quad \text{and} \quad \Lambda(\tau) = -2\bigg(\text{log}(L_{0}(\hat{\mu}_{0}))-\text{log}(L_{\tau}(\hat{\mu}_{1},\hat{\mu}_{2}))\bigg), \label{eq:LambdaTest}
\end{equation}
where \(\hat{\mu}_{0}=\hat{\beta}_{0}\) is the maximum likelihood estimator (MLE) for time series mean under the null hypothesis and \(L_{0}(\hat{\mu}_{0})\) is the corresponding model likelihood. \(L_{\tau}(\hat{\mu}_{1},\hat{\mu}_{2})\) is the likelihood function under the alternative hypothesis, where \(\hat{\mu}_{1}=\hat{\beta}_{0}\) and \(\hat{\mu}_{2}=\hat{\beta}_{0}+\hat{\beta}_{1}\) are the MLEs for the means of the two segments separated by changepoint \(t=\tau\). To address the issue of volatility in the test statistic near the boundaries, a cropped \(\Lambda_{\text{max}}\) is commonly used. Instead of maximizing over the entire range \(1<\tau\leq N\), the admissible changepoint locations are truncated between \(\ell\) (close to 1) and \(h\) (close to N). \citet{robbins2011mean} show that
\begin{equation}
    \Lambda_{\text{max}}\overset{D}{\to} \sup_{\ell\leq\tau\leq h}\frac{B^{2}(t)}{t(1-t)},  \label{eq:LambdaDist}
\end{equation}
where \(\{ B(t) \}_{t=0}^{1}\) denotes a standard Brownian bridge process and \(\overset{D}{\to}\) signifies convergence in distribution, and provide the p-value calculation for this test.

To compute the test statistic \(\Lambda_{\text{max}}\), one maximizes \(\Lambda(\tau)\) over all admissible changepoints \(\ell\leq\tau\leq h\). This procedure involves fitting alternative models a total of \(h-\ell+1\) times. In cases with shorter time series, the number of candidate changepoints is limited, making the hypothesis testing procedure efficient. However, for longer time series with large sample sizes, the candidate set of changepoints expands, necessitating numerous model fittings and significantly increasing computational costs. In a recent study by \citet{li2026MTMcpt}, under the marginalized transition model framework, they proposed a genetic algorithm approach for the AMOC problem. Since the \(\text{log}(L_{0}(\hat{\mu}_{0}))\) part in Equation \eqref{eq:LambdaTest} remains constant for every admissible \(\tau\), the objective function only fits the model under the null hypothesis once and inputs the value as an argument \texttt{logL0}. This approach intelligently searches for the optimal solution that maximizes \(\Lambda_{\tau}\) without exhaustively fitting alternative models for every candidate changepoint. The optimized \(\Lambda_{\text{max}}\) value is then compared with the critical value from the limiting distribution in Equation \eqref{eq:LambdaDist} to facilitate the decision-making in hypothesis testing.

We designed the chromosome representation and the objective function to maximize \(\Lambda(\tau)\) under the alternative hypothesis. The related R function used here (recalling that our default optimization is to minimize the objective function), is defined as follows:

\begin{verbatim}
lambda_tau <- function(chromosome, plen = 0, Xt, XMat, logL0) {
  N <- length(Xt)
  m <- chromosome[1]
  tau <- chromosome[2]
  if (tau < floor(0.05 * N) | tau > ceiling(0.95 * N)) {
    Lambdatau <- -2 * (logL0 + 1e6)
  } else {
    cpt <- c(rep(0, tau), rep(1, N - tau))
    DesignX <- cbind(XMat, cpt)
    fit <- arima(Xt, order = c(1, 0, 0), xreg = DesignX, include.mean = FALSE)
    logLa <- fit$loglik
    Lambdatau <- -2 * (logL0 - logLa)
  }
  return(-Lambdatau)
}
\end{verbatim}

\noindent The \CRANpkg{changepointGA} package offers users the flexibility to define custom genetic operators. The functions \texttt{amoc\_population}, \texttt{amoc\_selection}, \texttt{amoc\_crossover}, and \texttt{amoc\_mutation} from our \CRANpkg{changepointGA} can be treated as the custom genetic operators and be integrated into arguments in \texttt{cptga} and \texttt{cptgaisl}. Detailed explanations and illustrations of these custom genetic operator functions can be found in the function help documentation or the supplementary materials of \citet{li2026MTMcpt}. With the estimated \(\text{log}L_{0}(\hat{\mu}_{0})\) under the null hypothesis, we can run the genetic algorithm to approximately but efficiently find the \(\hat{\tau}\) to maximize \(\Lambda(\tau)\).

\begin{verbatim}
N <- 1000
XMatT <- matrix(1, nrow = N, ncol = 1)
Xt <- ts_sim(
  Ts = N,  beta = 0.5, XMat = XMatT, sigma = 1, phi = 0.5, theta = NULL, Delta = 2,
  CpLoc = 500, seed = 1234
)
DesignXEst <- XMatT
fit0 <- arima(Xt, order = c(1, 0, 0), xreg = DesignXEst, include.mean = FALSE)
logL0 <- fit0$loglik
res.changepointGA <- cptga(
  ObjFunc = lambda_tau, N = N, popSize = 200,
  pcrossover = 0.95, pmutation = 0.3, pchangepoint = 10 / N,
  popInitialize = "amoc_population",
  selection = "amoc_selection",
  crossover = "amoc_crossover",
  mutation = "amoc_mutation",
  Xt = Xt, XMat = XMatT, logL0 = logL0
)
summary(res.changepointGA)
\end{verbatim}

\begin{verbatim}
#> ###############################################
#> #         Changepoint Detection via GA        #
#> ###############################################
#>    Settings: 
#>    Population size         =  200 
#>    Number of generations   =  5057 
#>    Crossover probability   =  0.95 
#>    Mutation probability    =  0.3 
#>    Changepoint probability =  0.01 
#>    minDist                 =  1 
#>    Task mode               =  cp 
#>    Parallel Usage          =  FALSE 
#>    Seed                    =  NULL 
#> 
#> ##### GA results ##### 
#>    Optimal Fitness value = -124.5826 
#>    Optimal Solution: 
#>         Number of Changepoints =  1 
#>         Changepoints Locations =  486
\end{verbatim}

\begin{verbatim}
res.Island.changepointGA <- cptgaisl(
  ObjFunc = lambda_tau, N = N, popSize = 200, numIslands = 5,
  pcrossover = 0.95, pmutation = 0.3, pchangepoint = 10 / N,
  popInitialize = "amoc_population",
  selection = "amoc_selection",
  crossover = "amoc_crossover",
  mutation = "amoc_mutation",
  Xt = Xt, XMat = XMatT, logL0 = logL0
)
summary(res.Island.changepointGA)
\end{verbatim}

\begin{verbatim}
#> ###############################################
#> #  Changepoint Detection via Island Model GA  #
#> ###############################################
#>    Settings: 
#>    Population size         =  200 
#>    Number of Island        =  5 
#>    Island size             =  40 
#>    Number of generations   =  5000 
#>    Number of migrations    =  100 
#>    Crossover probability   =  0.95 
#>    Mutation probability    =  0.3 
#>    Changepoint probability =  0.01 
#>    minDist                 =  1 
#>    Task mode               =  cp 
#>    Parallel Usage          =  FALSE 
#>    Seed                    =  NULL 
#> 
#> ##### Island Model GA results ##### 
#>    Optimal Fitness value = -124.5826 
#>    Optimal Solution: 
#>         Number of Changepoints =  1 
#>         Changepoints Locations =  486
\end{verbatim}

\begin{verbatim}
fit.Island.changepointGA <- res.Island.changepointGA@overbestfit
tau.Island.changepointGA <- res.Island.changepointGA@overbestchrom
pvaluecalc <- function(x, ell, h) {
  res <- sqrt(x * exp(-x) / 2 * pi) * (1 - 1 / x) * log(((1 - ell) * h) / (ell * (1 - h))) + 4 / x
  return(res)
}
AMOCpvalue <- pvaluecalc(x = -fit.Island.changepointGA, ell = 0.05, h = 0.95)
AMOCpvalue
\end{verbatim}

\begin{verbatim}
#> [1] 0.0321072
\end{verbatim}

From the analysis above, both the \texttt{cptga} and \texttt{cptgaisl} functions identified a single changepoint at \(\hat{\tau}=\text{486}\) with \(\Lambda_{\text{max}}=\text{124.583}\). This corresponds to a p-value of 0.032 (less than the significance level of \(\alpha=0.05\)) via the calculation method for likelihood ratio test in \citet{robbins2011mean}. Therefore, we conclude that the identified changepoint at \(\hat{\tau}=\text{486}\) introduces a significant mean structural change.

\bibliography{RJreferences.bib}

\address{%
Mo Li\\
University of Louisiana at Lafayette\\%
Department of Mathematics\\ 217 Maxim Doucet Hall, Lafayette, LA 70504,\\ United States of America\\
\href{mailto:mo.li@louisiana.edu}{\nolinkurl{mo.li@louisiana.edu}}%
}

\address{%
QiQi Lu\\
Virginia Commonwealth University\\%
Department of Statistical Sciences and Operations Research\\ PO Box 843083, 1015 Floyd Ave., Richmond, VA 23284,\\ United States of America\\
\href{mailto:qlu2@vcu.edu}{\nolinkurl{qlu2@vcu.edu}}%
}

\end{article}

\end{document}